\documentstyle[aps,pra,multicol,graphicx]{revtex}

\begin{document}
\draft
%
\hoffset= -2.5mm

\title{Transport, Noise, and Conservation in the
Electron Gas:\\
How to Build a Credible Mesoscopic Theory}

\author
{Frederick Green}

\address{Centre for Quantum Computer Technology, School of Physics, \\
The University of New South Wales, Sydney, NSW, Australia,\\
and
Department of Theoretical Physics,
Research School of Physical Sciences and Engineering, \\
The Australian National University,
Canberra, ACT, Australia.}

\author{Mukunda P. Das}

\address{Department of Theoretical Physics,
Research School of Physical Sciences and Engineering, \\
The Australian National University,
Canberra, ACT, Australia.}

\maketitle

\begin{abstract}
Electron transport in metallic systems is governed by
four key principles of Fermi-liquid physics:
(i) degeneracy, (ii) charge conservation,
(iii) screening of the Coulomb potential, and (iv) scattering.
They determine the character of metallic conduction and noise
at mesoscopic scales, both near equilibrium and far from it.
Their interplay is described by {\em kinetic theory},
the serious method of choice for characterizing
such phenomena. We review microscopic kinetics for
mesoscopic noise, and in particular its
natural incorporation of the physics of Fermi liquids.
Kinetic theory provides a strictly conservative,
highly detailed description of current fluctuations
in quantum point contacts. It
leads to some surprising noise predictions.
These show the power of a model that respects
the microscopic conservation laws.
Models that fail in this respect are incorrect.
\end{abstract}


\begin{multicols}{2}

\section*{1. Introduction}

\subsection*{1.1 History}

Sustained, vigorous progress marks 100 years of thought on
a cornerstone of modern electronics: the physics of
metallic charge transport.
Its flowering
began with the classical insights of Boltzmann in the 19th
century and of Drude and Einstein early in the last one.
In the 1930s Sommerfeld and Bloch instituted the quantum
description of bulk metallic conduction
\cite{hoddeson}.
That development
culminated in the microscopically robust {\em Fermi-liquid picture}
\cite{nozieres,pinoz,abri},
proposed by Landau and Silin late in the 1950s.
\footnote{
It is important to keep in mind, for the rest of this paper,
that standard Fermi-liquid theory
was never conceived as a theory of the bulk. It has
{\em always} addressed metallic transport at {\em all} scales,
including the mesoscopic one. Indeed, its pedigree derives from
extreme quantum many-body problems,
such as nuclear matter and liquid $^3$He
\cite{pinoz}.
}

Fresh horizons have now opened up through the
vision of mesoscopic transport as quantum-coherent
transmission. This important innovation is credited to
Landauer's foresight
\cite{ldr57},
and has been deepened and extended since then by
Beenakker, B\"uttiker, Imry, and many others
\cite{bvh,buett,marlan,imry,imry2,ferry,datta,djb,blbu}.
Its achievements have been impressive.

As well as their novel emphasis on coherent scattering, the
modern theories of conduction advocate a second major shift,
one that is logically independent of the mechanism
for transport (quantum-coherent or otherwise).
This is claimed to solve
the subtle problems of {\em open boundary conditions}
\cite{frensley,sols,fenton,wims},
central to conduction in a real mesoscopic system
\cite{imry}.
It is intended to supplant
the long-dominant picture of charge flow as {\em drift}.

In drift, the current is the collective average response to
which every carrier contributes. It is the effect of an external cause,
the applied voltage.
Over against drift, the new mesoscopic transport
revisits the notion of charge flow as purely a kind of {\em diffusion}.
Here, it is the current that is regarded as externally supplied
\cite{imry}.
Introduction of current into the system sets up a virtual density
imbalance between the carrier reservoirs interconnected by the
transmissive device.
The observed voltage drop is merely a by-product of that
virtual imbalance.

Figure 1 illustrates the two viewpoints; diffusion and drift
are seen in their seemingly contrasting physical roles. 
This simple shift of perspective, from drift to diffusion,
has been extraordinarily successful in predicting
mesoscopic transport phenomena.
These have been carefully documented and explained
\cite{imry,imry2,ferry,datta,djb,blbu}.

The explanatory simplicity and consequent attractiveness of
diffusive phenomenologies does not mean, however,
that microscopically based analyses
of transport and noise have become redundant.
Microscopic methods, built upon the legacy that runs
from Boltzmann to Landau, are pursued with unabating vigor
\cite{bk,frensley,sols,fenton,wims,nvk,sw1,sw2,kormay,gillespie,gdi,gdii,ajp,fnl,kk}.
That said, phenomenological simplicity has never been
a foolproof guide to theoretical depth and correctness.
The situation in mesoscopic physics is no different.

\end{multicols}

\begin{minipage}{16cm}

\begin{figure}[h]

%
\begin{center}
\includegraphics[width=0.65\textwidth]{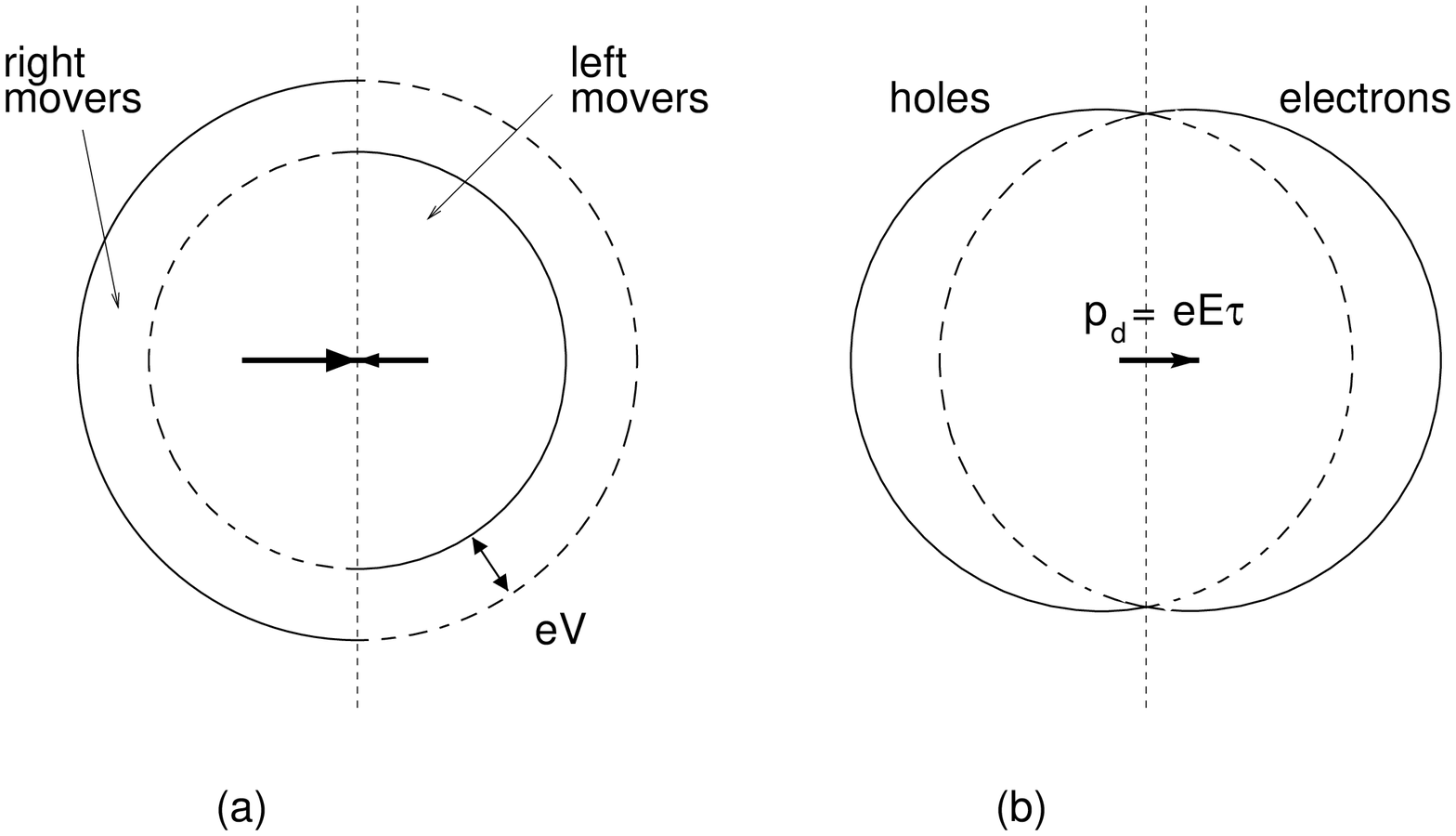}
\label{f1} 
\vspace{0.5cm}
\end{center}

{FIG. 1.
Diffusive and drift concepts of mesoscopic transport, compared.
(a) Diffusion. An applied electromotive
force $eV$ defines a mismatch between the quasi-Fermi energies
of degenerate electrons at the source and drain.
Only ``right movers'' at the higher-energy source
lead contribute to current flow; only ``left movers''
at the lower-energy drain lead contribute to the current counterflow.
The physical current is phenomenologically identified
with their difference. This pseudodiffusive current ``generates''
$eV$ if and only if one additionally assumes the
validity of Einstein's relation between diffusion and conductance.
{\em All carriers are at equilibrium; their role in transport is
passive}. There is no electron-hole symmetry
in pseudodiffusive transport
\cite{fenton}.
(b) Drift. All of the carriers that fill states in the Fermi
sea feel, and respond to, the external driving force $eE$.
Each gains average momentum $p_d = eE\tau$ by accelerating
ballistically during a mean time $\tau$ before rescattering.
The flux of carriers in deeper-lying filled states is canceled by
opposing filled states. Only those electron states {\em kinematically
matched} to holes, within a shell of thickness $p_d v_{\rm F}$
at the Fermi surface, contribute to the physical (drift) current.
{\em The volume of the Fermi sea remains invariant regardless of} $eE$.
The volume is rigidly fixed by the {\em equilibrium} Fermi energy.
There is automatic electron-hole symmetry in drift transport
\cite{fenton}.
}
\end{figure}
%
\end{minipage}
\vspace{0.5cm}

\begin{multicols}{2}

For a mesoscopic theory's credibility, only two questions count:

\begin{itemize}
\item
Does the theory fully
respect {\em all} of the essential physics of the interacting electron gas
\cite{pinoz}?
\item
If not, {\em why} not?
(Some discussion of this is in References
\onlinecite{gdi,upon,ithaca}.)
\end{itemize}

\noindent
Our goal is straightforward. We restate, and elaborate,
a plain theoretical fact. If a noise model is
truly microscopic -- faithful to the long-established
and completely orthodox procedures of kinetics
and electron-gas theory
\cite{pinoz,bk,nvk}
-- then it must, and does, produce reliable predictions
at mesoscopic scales. These may be quite surprising.

Microscopically based descriptions, for instance kinetic
ones, outstrip the scope of low-field phenomenologies
to access the strongly nonequilibrium regime.
Equally important is the fact that only a reliable microscopic
foundation can support the well controlled approximations
that are always needed to turn a generic theory
into a powerful, practical design tool for novel electronics. 

The heart of any kinetic approach is conservation.
Microscopic conservation implies that
diffusion and drift manifest as complementary
but {\em interlocking} effects in the physics.
They are in no sense mutually exclusive.
This crucial point needs a closer look.

\subsection*{1.2 Drift or Diffusion?}

Before setting out the plan of our paper, we briefly address
the folklore that transmissive-diffusive models
are more ``physical'' than (and somehow superior to)
wholly kinetic descriptions of mesoscopic transport.
For {\em uniform} systems, there is a formal congruence
between ``pure'' drift and ``pure'' diffusion.
They connect via the Einstein relation
\cite{kittel,vanvliet}
which links $\sigma$, the low-field conductivity
of a metal, to $D$, its equilibrium diffusion constant:

\begin{equation}
\sigma \equiv e^2 D {{\partial n}\over {\partial \mu}}
\label{Eq_0}
\end{equation}

\noindent
at carrier density $n$ and chemical potential $\mu$.
Substitution of diffusion for (weak-field) conductance is
justified when the system's shortest scattering mean free path
is much less than its length. However, it is claimed
that this clearly semiclassical Ansatz can be extended
even to quantum-coherent mesoscopics
\cite{datta}.

The conductivity quantifies the coarse-grained
{\em single-particle} current response; $\sigma$ is accessible
through the current-voltage characteristic.
The diffusion constant is
a fine-grained {\em two-body} response, and its
structure is intimately tied to current fluctuations;
$D$ too is observable, for example via time-of-flight
methods that are essentially two-point correlation measurements
\cite{nougier}.
Equation (\ref{Eq_0}) clearly shows that diffusion
and drift go hand-in-hand; it is not an either-or situation.

Einstein's relation between conductivity and diffusion
brings to the fore a central theme, namely
the {\em underlying unity of transport and fluctuations} (noise).
This unity, which is fundamentally microscopic,
is embodied in the fluctuation-dissipation theorem
(the Einstein relation is a special case).
It establishes the proportionality of dissipative transport
to the fluctuations inherent in the structure.
Such a theorem can never be proved heuristically
\cite{upon}.

This is the crucial point. Diffusive phenomenologies
are forced to invoke the fluctuation-dissipation
theorem as an external assumption. It is their {\em only means} to
justify, in an intuitive way, the linear current-voltage
characteristic on which they absolutely rely.
A transport model that chooses to favor
diffusion, merely for intuitive reasons,
denies the core {\em microscopic unity}
of noise and conductance. The fluctuation-dissipation
relation is then no longer a prescriptive, first-principles
constraint on the possible physics of the problem.
Instead it is reduced to a
highly compliant, imaginative guiding ``rule'';
one that can be molded to any set of favorite preconceptions.

For noise, diffusive (or, more accurately, pseudodiffusive)
descriptions invariably take this linear theorem on faith.
This is so that the current-current correlator
can be adjusted, by hand, to force it to fit the conductance.
Such maneuvers are necessary only because,
quite unlike microscopic theories (the Kubo formalism
\cite{kubo}
is a good example), diffusive phenomenologies cannot
express -- and thus compute -- their
correlators from first principles.

The transmissive-diffusive models lack
a formal basis for deriving the
fluctuation-dissipation theorem
\cite{upon,ithaca}.
That result is provable only
within a microscopic description, embedded
in {\em statistical mechanics}
\cite{kubo},
or else in {\em kinetic theory}
\cite{nvk}.
Models of the Landauer-B\"uttiker-Imry class
share little, if any, of that essential machinery.

Few mesoscopic systems are truly homogeneous on the
length scale over which transport unfolds.
Generally, the mode of electron transfer through
a nonuniform channel is not by real-space diffusion alone,
or by drift alone (that is: diffusion in velocity space).
Actual mesoscopic transport is some combination of drift and
diffusion, physically conditioned by the nonuniformities
specific to the system.
For instance, the electron gas in a III-V heterojunction quantum well
\cite{vinter}
is extremely nonuniform in the direction of crystal growth,
normal to the plane of conduction.
(This also leads to strong quantum confinement and to marked suppression
of the fluctuations for the two-dimensional carriers
\cite{gdii}.)
To insist that one transport mode is absolutely dominant
is to risk distorting the real physics. 

Only a description that treats diffusion and drift on an equal
footing, favoring neither one process nor the other
{\em ad hoc}, is able to span in a unified fashion
the complete range of transport and noise physics.
Orthodox kinetic theory
\cite{nvk},
coupled with precise microscopic knowledge of the electron gas
\cite{pinoz},
provides exactly that description. It
accommodates both {\em nonuniform-field effects} and
{\em nonequilibrium response}.

Finally we recall that weak-field approaches of the
transmissive-diffusive kind tend to
assume that the metallic electron gas is well described as
a group of free, noninteracting fermions subject only to
elastic scattering
\cite{imry}.
It means that self-consistent collective
screening -- ever preeminent in the electron gas -- is
regarded as a secondary perturbation
(if, in fact, it is believed to matter at all).
Such theories are not set up to describe strongly nonuniform Coulomb
correlations
\cite{gdii},
any more than they can treat the strongly
nonequilibrium domain where dissipative {\em inelastic}
collisions rule explicitly
\cite{fnl}.

\subsection*{1.3 Issues for Review}

To venture into the important regimes of
high-field transport and Coulomb correlations,
much more is demanded of a
mesoscopic theory than is deliverable by current descriptions
\cite{imry,imry2,ferry,datta,djb,blbu}.
Among the sea of literature, it is still
unusual to find theories of metallic conduction that
explicitly adopt clear and firmly validated microscopic methods.
At and beyond the low-field limit, a small but growing
number of kinetic approaches exists
\cite{sw1,sw2,kormay,gdi,gdii,ajp,fnl,kk},
designed to answer the often-stated need
\cite{blbu}
for new mesoscopic approaches, especially
away from equilibrium.

For novel technologies, if not for the sake of fundamental physics alone,
closure of this knowledge gap is a significant task.
Our own endeavors are detailed in
Refs.
\onlinecite{gdi,gdii,ajp,fnl,upon,ithaca,balan,cmt25}.
The present work is an up-to-date survey of that research.

In Section 2 we briefly introduce the two primary results of our
exactly conserving kinetics: (i) {\em thermal scaling}
and (ii) {\em Coulomb-induced suppression}
of nonequilibrium fluctuations in a mesoscopic metallic conductor. 
We discuss their physical meaning, and their place in
a coherent understanding of mesoscopic noise.
While these core concepts are easy to state, their
formal basis requires elaboration. This is given in
Sec. 3; we cover the roles of {\em degeneracy},
{\em conservation},  and {\em screening}. For that we draw
on the Landau-Silin equation of motion
\cite{pinoz},
itself an extension of Boltzmann transport to charged Fermi liquids.
In Sec. 4 we turn to a significant application:
nonequilibrium ballistic fluctuations in one dimension.
Our strictly conservative kinetic model leads to some
surprises. We state our conclusions in Sec. 5.

\section*{2. Physics of the Boundary Conditions}

\subsection*{2.1 Ground Rules for Nonequilibrium Transport}

In this Section we discuss two elementary, and indispensable,
boundary constraints on an externally driven open conductor. They are
\cite{gdi}:

\begin{itemize}

\item
{\em Global charge neutrality} over the conductor, its interfaces,
and the connected source and drain reservoirs. Gauss' theorem
implies {\em unconditional} global charge neutrality; that is,
a neutrality that is absolutely independent of the dynamics
within the active body of the device.

\item
{\em Local thermodynamic equilibrium} in each source and drain lead
interfacing with the device.
Energetic stability means that each of these local reservoir
equilibria is also {\em unconditional} and independent of internal
dynamics.

\end{itemize}

\noindent
Both of these are universally understood as crucial for
transport in open systems, yet their microscopic consequences
seem not to be understood as well. We expand on them.

Figure 2 shows a generic two-terminal situation. The device
is in intimate electrical contact with its two stabilizing reservoirs
while a closed loop, incorporating an ideal generator, sustains a
controlled current between drain and source.
The system attempts to relax via net charge displacement
across the source and drain.
The induced potential -- Landauer's resistivity dipole
\cite{ldr57}
-- is the response.
Equivalently, a closed loop with an ideal battery in series
with the structure can be created, exerting a
controlled electromotive force (EMF) {\em locally}
across the active region
\cite{wims}.
The response is the carrier flux induced in the loop.

\end{multicols}


\begin{figure}[h]

%
\begin{center}
\includegraphics[width=0.40\textwidth]{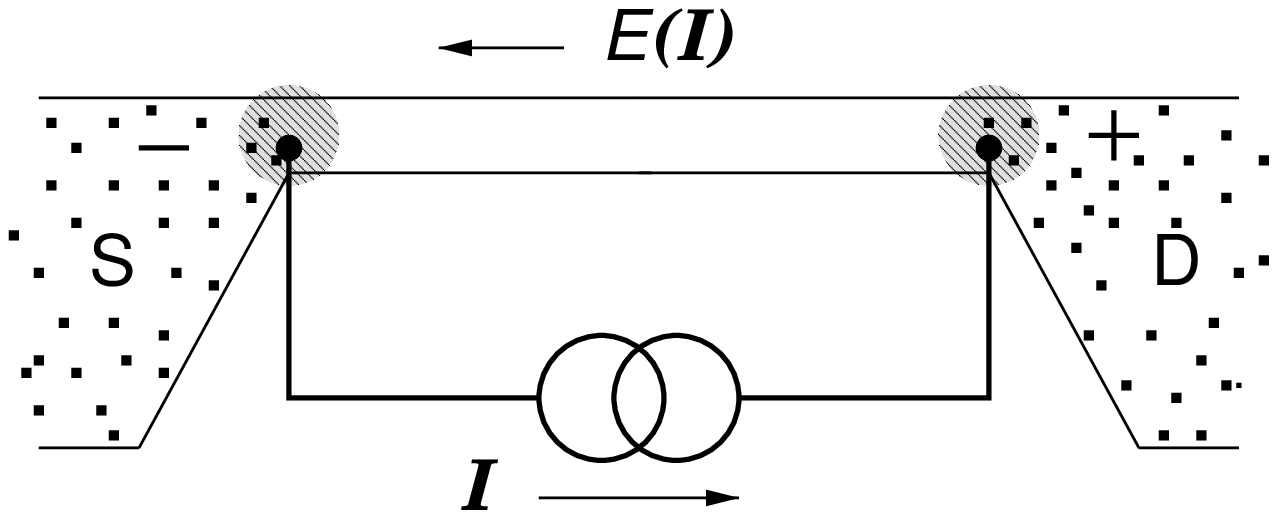}
\label{f2} 
\vspace{0.5cm}
\end{center}

{FIG. 2. An idealized mesoscopic conductor. Its diffusive leads
(S, D) are in unconditional equilibrium. A paired source and sink
of current $I$ at the boundaries explicitly drives the transport.
Local charge clouds (shaded) are induced by the active influx and
efflux of $I$. These are regions of vigorous and dynamic competition
among the current-driven excitation of carriers, their elastic
{\em and inelastic} dissipative relaxation, and strong Coulomb
screening from the stabilizing lead reservoirs. Together, these
competing effects establish the self-consistent dipole potential
$E(I)L$ across distance $L$ between drain and source;
that potential is the electromotive
force in the driven system.
}

\end{figure}
\vspace{0.5cm}

\begin{multicols}{2}

\subsection*{2.2 Charge Conservation for Open Systems}

In either of the two scenarios (fixed current or fixed EMF),
the specific action of the external flux sources and sinks
ensures that electronic transport through
the open system conserves charge
\cite{sols}.
Entry and exit of the current in a mesoscopic conductor
cannot be treated by vague appeals to asymptotic equilibrium
\cite{imry}.
That is because entry and exit of the current is
always a dynamic nonequilibrium process.

To guarantee global gauge invariance, all sources and sinks {\em must
be considered explicitly} as part of the dynamical description
of the transport
\cite{sols}.
If not, the price is clear. It is the loss of charge conservation,
and an ill-conceived model.

Under all circumstances, the current sources and sinks,
and the EMF, are {\em localized} inside
a finite volume that also encloses the conductor
\cite{fenton,wims}.
This, like global neutrality,
is a necessary consequence of gauge invariance
\cite{sols}.
Outside the active volume, the undisturbed electron population
within each lead (stabilized by its compensating positive background)
always remains charge-neutral and pins the local Fermi level
within that lead.
It means that the nonequilibrium carriers in the active,
and finite, conducting channel have to reconnect smoothly
to the {\em invariant} local equilibrium state
beyond the interfaces
\cite{gdii}.

The reservoir equilibria (each one locally proper to its lead)
remain totally unaffected by the transport dynamics. None of the local
density-dependent quantities within the leads,
including their fluctuations, ever changes. None ever responds to
the possibly extreme conditions in the driven device.
This proves to be a formidable constraint on what can happen
inside.

\subsection*{2.3 Constraint on the Total Carrier Number}

Let $f_{\bf k}({\bf r},t)$ be the time-dependent electron distribution
for wave vector ${\bf k}$, at point ${\bf r}$ in the active region.
Spin and subband labels are understood (for simplicity we
take only twofold spin degeneracy). From the microscopic object
$f_{\bf k}({\bf r},t)$, all the physical one-body properties
can be calculated, such as the mean electron density $n({\bf r},t)$
and the current density ${\bf J}({\bf r},t)$.

If $N$ is the total number of carriers
within the region, of volume $\Omega$ say,
\footnote{
It is absolutely essential to include the interface regions
(the buffer zones where all the fringing fields are extinguished
by screening) as part of the active volume of the driven device.
}
then a sum of local momentum states over the entire 
active region, of dimension $\nu = 1, 2$, or 3, leads to

\begin{equation}
\int_{\Omega} d{\bf r} \int {2d{\bf k}\over (2\pi)^{\nu}}
f_{\bf k}({\bf r},t)
= N =
\int_{\Omega} d{\bf r} \int {2d{\bf k}\over (2\pi)^{\nu}}
f^{\rm eq}_{\bf k}({\bf r})
\label{Eq_1}
\end{equation}

\noindent
where $f^{\rm eq}_{\bf k}$ is the equilibrium distribution.
The mean total carrier number is constant and
remains fully compensated by the
nonparticipating positive background,
integrated over $\Omega$.

Equation (\ref{Eq_1}) makes a straightforward statement.
Gauss' theorem implies -- unconditionally -- that
the device remains overall neutral at any driving field.
This is true if and only if the inner active region is
efficiently screened from the macroscopic leads by the
electron gas at the interfaces
\cite{kk}.
Mean-field screening (Poisson's equation) thus
ensures the leads' (local) neutrality at all times,
while the asymptotic equilibrium of each lead ensures that the
total volume $\Omega$, where nonequilibrium processes take place,
is fixed and finite.

Whether in equilibrium or not, we have the principle that

\begin{itemize}
\item
{\em Within the active mesoscopic structure,
the mean total number of mobile carriers is invariant}.
\end{itemize}

\noindent
Belying its almost self-evident nature, this rule has profound
implications for the fluctuations of the nonequilibrium state.

\subsection*{2.4 Constraint on Total Fluctuation Strength}

Random external perturbations give rise to a persistent fluctuation
background. This displaces the instantaneous distribution
$f_{\bf k}({\bf r},t)$ from its steady-state ensemble average.
The {\em same} external stochastic processes
\footnote{
Examples are {\em quasicontinuous} energy exchange
with phonons in the thermal bath of the lattice
(generating thermal noise),
and {\em discrete} Poissonian injection/extraction of carriers by
the external sources/sinks of current (generating shot noise).
}
act on the channel both at equilibrium and when it is driven
by an injected current (or by a battery-generated EMF).

Let $\Delta N \equiv k_{\rm B}T\partial N/\partial \mu$
be the mean-square thermal number fluctuation.
Then Gauss' theorem acts as a constraint on Eq. (\ref{Eq_1})
for $N$, taking note that the latter is (potentially)
a dynamical quantity.
As a result, global neutrality enforces a {\em fluctuation}
counterpart to the sum rule of Eq. (\ref{Eq_1}). This involves,
in the one relation, the mean-square thermal fluctuation $\Delta f(t)$
of the single-particle distribution $f(t)$,
and its basic equilibrium form $\Delta f^{\rm eq}$:

\begin{equation}
\sum_{\alpha} \Delta f_{\alpha}(t)
= \Delta N =
\sum_{\alpha} \Delta f^{\rm eq}_{\alpha}.
\label{Eq_2}
\end{equation}

\noindent
For brevity, we have condensed the  notation.
We now use the composite state-labels
$\alpha \equiv ({\bf k}, {\bf r})$,
$\alpha' \equiv ({\bf k'}, {\bf r'})$ and so on,
while the generalized sum (with spin degeneracy)
is defined by

\[
\sum_{\alpha} {\cdot \cdot \cdot} {~~}\equiv
\sum_{\bf r} \Omega({\bf r}) \sum_{\bf k} {2\over \Omega({\bf r})}
{~} {\cdot \cdot \cdot}
{~~}\equiv
\int_{\Omega} d{\bf r} \int {2d{\bf k}\over (2\pi)^{\nu}}
{~~} {\cdot \cdot \cdot}
\]

\noindent
in which the working volume $\Omega$ is subdivided, in a
standard way, into sufficiently small local cells $\Omega({\bf r})$ that
are still large compared to the particle volume
$n^{-1}$.
(The unit cell volume in reciprocal space becomes $\Omega({\bf r})^{-1}$.)
The equilibrium fluctuation
$\Delta f^{\rm eq}_{\alpha}$ is determined
from standard statistical mechanics:
\footnote{The microscopic structure of ${\Delta f^{\rm eq}}$
is richer than its simple statistical mechanics definition suggests.
It is better to recall its kinetic
origin as a quantum-correlated {\em electron-hole excitation}
taken in its long-wavelength static limit
\cite{pinoz}:
\\
\[
{{\Delta f^{\rm eq}_{\bf k}}\over k_{\rm B}T}
=
-\lim_{{\bf q} \to 0}
{\left[
\lim_{\omega \to 0}
{\left(
{ {f^{\rm eq}_{{\bf k} - {\bf q}/2}
 - f^{\rm eq}_{{\bf k} + {\bf q}/2}}
  \over
  {\hbar\omega - \varepsilon_{{\bf k} + {\bf q}/2}
               + \varepsilon_{{\bf k} - {\bf q}/2}} }
\right)}
\right]}
\]
\\
\noindent
for particle band energy $\varepsilon_{\bf k}$.
}

\[
\Delta f^{\rm eq}_{\alpha} \equiv k_{\rm B}T
{\partial f^{\rm eq}_{\alpha}\over
{\partial \varepsilon_{\rm F}({\bf r})}}
= f^{\rm eq}_{\alpha}(1 - f^{\rm eq}_{\alpha}).
\]

\noindent
Here the local electrochemical potential
$\varepsilon_{\rm F}({\bf r}) = \mu - U_0({\bf r})$,
basically the Fermi level of the local population,
is given by the global chemical potential $\mu$
offset by the mean-field (Hartree) potential $U_0({\bf r})$.

Equation (\ref{Eq_2}) is a rigorous, nonequilibrium,
kinetic-theoretical relation
\cite{gdii}.
It controls the physics of thermal fluctuations at
length scales greater than the metallic Fermi wavelength,
which is itself short (0.2--10 nm) compared to
mesoscopic device sizes (say 50--1000 nm).

\subsection*{2.5 Temperature Scaling}

Two outcomes flow from Eq. (\ref{Eq_2}).
The first is that even the {\em nonequilibrium}
thermal fluctuations in a degenerate conductor
necessarily scale with the thermal energy $k_{\rm B}T$,
whatever the value of the driving voltage.
For a specific illustration, see Fig. 3.
The closed microscopic form of the distribution
$\Delta f_{\alpha}(t)$ is given explicitly
in Sec. 3 below. For the moment we state a milder result,
the sum rule for the total fluctuation strength
in the degenerate limit:

\begin{mathletters}
\label{Eq_3}

\begin{equation}
\sum_{\alpha} \Delta f_{\alpha}(t)
= \sum_{\alpha} \Delta f^{\rm eq}_{\alpha}
\to k_{\rm B}T
\sum_{\bf r} \Omega({\bf r})
{\cal D}[\varepsilon_{\rm F}({\bf r})],
\label{Eq_3a}
\end{equation}

\noindent
in which the Fermi-Dirac form of $\Delta f^{\rm eq}$
is used to introduce the density of states ${\cal D}$:

\begin{eqnarray}
{2\over \Omega({\bf r})} \sum_{\bf k} \Delta f^{\rm eq}_{\alpha}
=&& 2k_{\rm B}T \sum_{\bf k}
{\left\{
{\delta(\varepsilon_{\alpha} - \varepsilon_{\rm F}({\bf r}))
\over \Omega({\bf r})}
\right\}}
\cr
{\left. \right.} \cr
&&\to k_{\rm B}T {\cal D}[\varepsilon_{\rm F}({\bf r})]
\label{Eq_3b}
\end{eqnarray}

\end{mathletters}

\noindent
where $\varepsilon_{\alpha}$ is the local band energy of a carrier.

There is an immediate corollary for
the {\em current autocorrelation function}, which
shapes the observable noise spectrum for the structure.
The thermal current correlations will scale with $\Delta f$.
Equation (\ref{Eq_3}) asserts that the thermal
contribution to noise must exhibit a strict proportionality to
the base temperature $T$, even well away from
the linear low-field regime
(where the Johnson-Nyquist formula itself
\cite{kittel}
enforces $T$-scaling).

A question arises naturally: How can this behavior be reconciled
with the appearance of shot noise, a thermally insensitive effect?
The kinetic-theoretical answer (which we justify, fully
and formally, in Sec. 3)
is uncompromising:

\begin{itemize}
\item
{\em There is no continuous transformation
(crossover) of thermal noise into shot noise}.
\end{itemize}

\noindent
As a purely nonthermal fluctuation effect, shot noise
can never satisfy the rigid sum rule expressed
in Eqs. (\ref{Eq_2}) and (\ref{Eq_3}).
Nor does it satisfy the fluctuation-dissipation theorem;
an in-depth analysis of this and other essential
distinctions between shot noise and thermal noise is given
by Gillespie
\cite{gillespie}.

Equation (\ref{Eq_3}), and Fig. 3, directly countermand
the Landauer-B\"uttiker account of shot noise as all of one piece
with thermal noise.
\cite{blbu};
thus they make a nontrivial statement. For a
kinetic-equation approach to shot noise see Refs.
\onlinecite{ajp,upon}.
For complete technical details of that approach, see Ref.
\onlinecite{9809339}.


\end{multicols}

\begin{figure}[h]

%
\begin{center}
\includegraphics[width=0.30\textwidth]{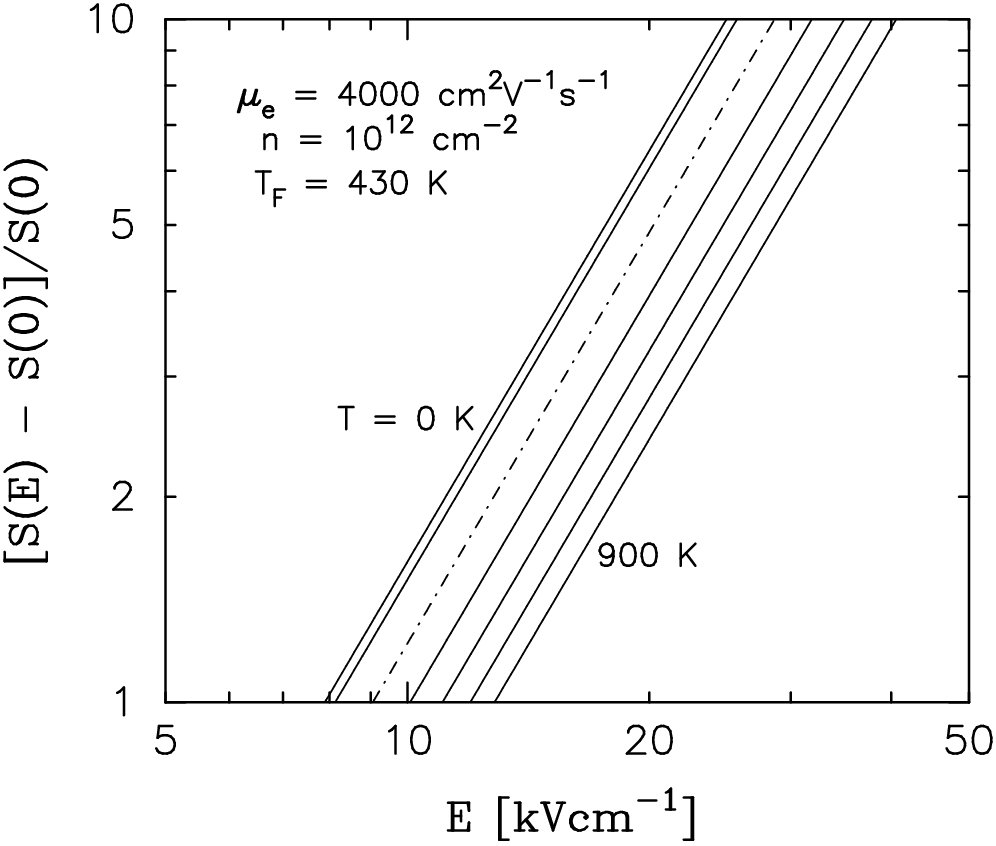}
\label{f3} 
%
\end{center}

{FIG. 3. Temperature scaling of degenerate hot-electron noise.
The nonequilibrium excess spectrum is for carriers confined
in an AlGaAs/InGaAs/GaAs heterojunction quantum well
at electron density $10^{12}{~}{\rm cm}^{-2}$
and mobility $4000{~}{\rm cm}^2{\rm V}^{-1}{\rm s}^{-1}$.
The hot-electron noise is plotted for fixed temperature
($T$ goes from 0 to 900 K in increments of 150 K),
as a function of applied electric
field. Normalization is to the
Johnson-Nyquist value $S(E=0) = 4G k_{\rm B}T$. In the
limit $T \to 0$ we have $S(E) \propto T/T_{\rm F}$
for Fermi temperature $T_{\rm F} = \varepsilon_{\rm F}/k_{\rm B}$.
Degeneracy forces the hot-electron noise to vanish with
temperature, so the ratio $[S(E) - S(0)]/S(0)$ is independent of $T$.
In the limit $T \gg T_{\rm F}$ the electrons are classical.
The excess noise is independent of $T$
so that $[S(E) - S(0)]/S(0) \ll 1$.
The dot-dashed line is for $T = 300$ K.
}

\end{figure}

\begin{multicols}{2}

\subsection*{2.6 Noise Suppression via Degeneracy and Inhomogeneity}

The second outcome of the microscopic theory leading
to Eq. (\ref{Eq_2}), as detailed in the next Section,
is that a mesoscopic conductor which is
strongly nonuniform manifests {\em Coulomb suppression
of charge fluctuations} below those of a uniform reference
medium, with otherwise identical transport characteristics
\cite{gdii}.

One other condition is essential for Coulomb suppression: 
carrier degeneracy.
Suppression is a unique effect of Fermi statistics,
acting in conjunction with spatial inhomogeneity
and Coulomb screening. It is not seen in a classical
electron gas, where Maxwell-Boltzmann statistics
leads to equipartition of the internal energy
\cite{kittel}.

The mechanism of suppression is as follows.
Degenerate carriers in a nonuniform channel experience
some degree of localization. They will lower their total
energy by a partial rearrangement, setting up a
self-consistent field to screen their large charging
energy, due to degeneracy and confinement.
The confining potential can be engineered by spatially
dependent doping, discontinuities in the band structure,
or a combination of both, as in most
III-V heterojunction quantum channels
\cite{vinter}.

Taking the latter as our example, let us
look for the effect of the large self-consistent
Coulomb energy on the fluctuations of the
two-dimensional electron gas (2DEG).
A channel with density $n_s$ in the plane of confinement
contains $\Omega n_s$ carriers in area $\Omega$:

\begin{equation}
N \equiv \Omega n_s = \Omega {\cal D}k_{\rm B}T
\ln {\{ 1 + \exp[(\mu - \varepsilon_0(n_s))/k_{\rm B}T] \}}.
\label{Eq_4}
\end{equation}

\noindent
Here ${\cal D} = m^*/\pi \hbar^2$ is the 2DEG density of states.
For simplicity we assume ground-state 
occupation only, at subband energy $\varepsilon_0(n_s)$.
The density dependence of $\varepsilon_0(n_s)$
reflects the strong Coulomb repulsion within
the 2DEG, confined in the quantum well perpendicular
to the channel.

Equation (\ref{Eq_4}) can be varied in two ways to arrive at the
charge-fluctuation strength over the channel. If the
internal potential is frozen,
$\varepsilon_0(n_s)$ remains at a {\em fixed value}.
With this variational restriction, the 2DEG form of
Eq. (\ref{Eq_2}) for the driven channel becomes
\cite{gdii}

\begin{eqnarray}
\sum_{\alpha} \Delta f_{\alpha}(t)
=&& \Delta N
= k_{\rm B}T
{\Biggl.
{{\partial N}\over {\partial \mu}}
\Biggr|}_{\varepsilon_0(n_s)}
\cr
{\left. \right.} \cr
=&& { {\Omega {\cal D}k_{\rm B}T}\over
{1 + \exp[(\varepsilon_0(n_s) - \mu)/k_{\rm B}T]} }.
\label{Eq_5}
\end{eqnarray}

\noindent
Lifting the restriction on the internal potential
now allows for the natural, self-consistent relaxation of
the local field due to the charge fluctuations. We do this
by including the negative-feedback term that comes from the
density dependence of $\varepsilon_0(n_s(\mu))$, present
on the right-hand side of Eq. (\ref{Eq_4}).
The self-screening of $\Delta f_{\alpha}$ then means that


\begin{mathletters}
\label{Eq_6}

\begin{eqnarray}
k_{\rm B}T {{\delta N}\over {\delta \mu}}
=&& \sum_{\alpha} {\widetilde \Delta}f^{\rm eq}_{\alpha}
= {\left( 1 - {{\delta \varepsilon_0}\over {\delta \mu}} \right)}
\sum_{\alpha} \Delta f^{\rm eq}_{\alpha}
\cr
{\left. \right.} \cr
=&& {\left( 1 - {1\over \Omega}{{\delta N}\over {\delta \mu}}
{{d \varepsilon_0}\over {d n_s}} \right)}
k_{\rm B}T{\Biggl.
{{\partial N}\over {\partial \mu}}
\Biggr|}_{\varepsilon_0(n_s)}
\label{Eq_6a}
\end{eqnarray}

\noindent
where ${\widetilde \Delta} f^{\rm eq}_{\alpha}$ is
the equilibrium distribution of fluctuations,
in the full presence of self-consistency.
Eq. (\ref{Eq_6a}) can be rearranged to
give a closed expression for the total number fluctuation

\begin{equation}
{{\widetilde \Delta} N} = k_{\rm B}T{{\delta N}\over {\delta \mu}}
= { {\Delta N}\over
{1 + {\displaystyle {\left( {{\Delta N}\over {\Omega k_{\rm B}T}} \right)}
                {d\varepsilon_0\over dn_s}}} },
\label{Eq_6b}
\end{equation}

\noindent
in complete analogy with the Thomas-Fermi screening
formula for the bulk electron gas
\cite{pinoz}.

Through the global-neutrality condition,
Gauss' theorem again leads straight to a dynamical sum rule
for the 2DEG fluctuations:

\begin{equation}
\sum_{\alpha} {\widetilde \Delta}f_{\alpha}(t)
= {\widetilde \Delta} N
= { {\Delta N}\over
{1 + {\displaystyle {\left( {{\Delta N}\over {\Omega k_{\rm B}T}} \right)}
                {d\varepsilon_0\over dn_s}}} },
\label{Eq_6c}
\end{equation}

\end{mathletters}

\noindent
where ${\widetilde \Delta} f_{\alpha}(t)$ denotes the
time-dependent mean-square distribution of
the fluctuations out of equilibrium, with full self-consistency.
Eq. (\ref{Eq_6c}), like Eq. (\ref{Eq_2}) before it,
is an exact relation with a rigorous kinetic-theoretical basis
\cite{gdii}.

\end{multicols}


\begin{figure}[h]

%
\begin{center}
\includegraphics[width=0.30\textwidth]{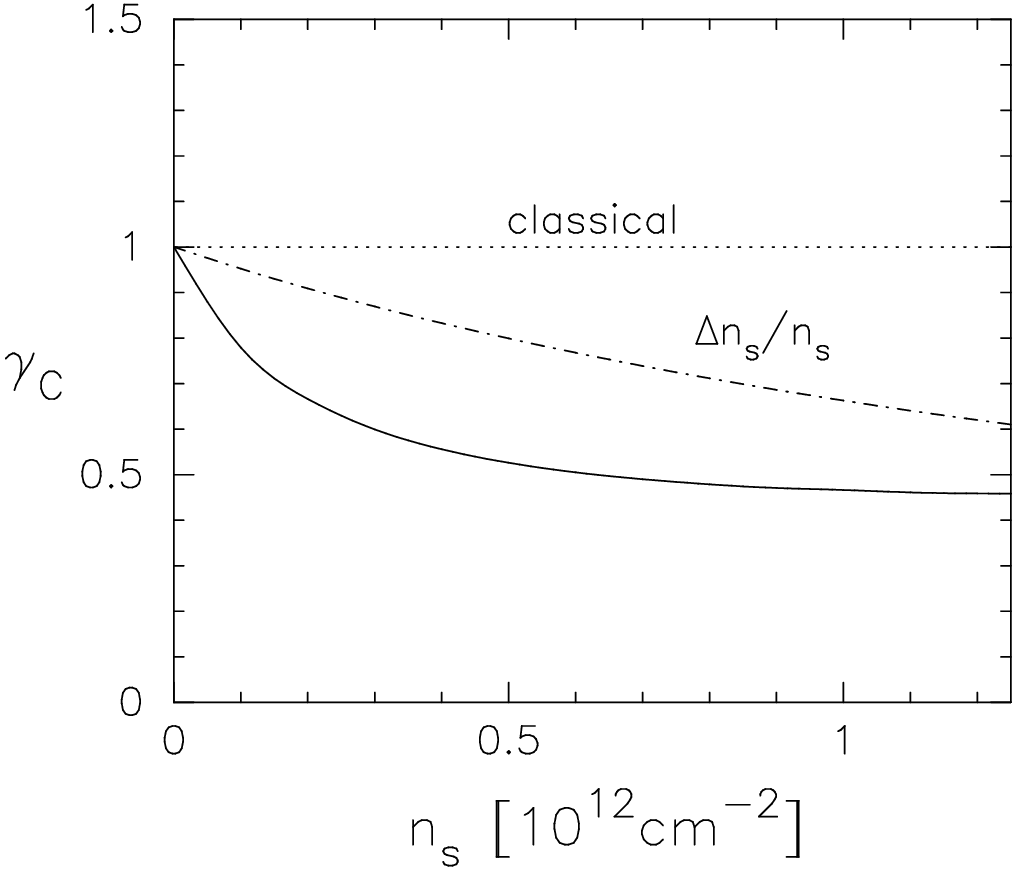}
\label{f4} 
\vspace{0.5cm}
\end{center}

{FIG. 4. Effect of inhomogeneous Coulomb screening on the
quantum-well confined electron population in an AlGaAs/InGaAs/GaAs
heterojunction, as a function of sheet electron density $n_s$.
Solid line: the suppression coefficient for degenerate carrier
fluctuations, $\gamma_{\rm C} \equiv {\widetilde \Delta} N/\Delta N$;
refer to Eq. (\ref{Eq_6c}) in the text. Dot-dashed line:
The unscreened (free-carrier) ratio $\Delta N/N$
of mean-square number fluctuations to mean carrier number.
This ratio measures the degeneracy of the
system; a smaller value means higher degeneracy. Both
$\Delta N/N$ and $\gamma_{\rm C}$ are intimately related to
the system's compressibility; see Eqs. (\ref{Eq_6.2}) and (\ref{Eq_6.3}).
Dotted line: in the classical limit both ratios
are unity. {\em When there is no degeneracy, there is no
inhomogeneous Coulomb suppression of the compressibility}.
}

\end{figure}

\begin{multicols}{2}

In Figure 4 we show the behavior of equilibrium fluctuations
in a pseudomorphic AlGaAs/InGaAs/GaAs heterojunction
at room temperature. Under normal operating
conditions, even without cryogenic cooling, the quantum
confined electron gas suppresses its thermal fluctuations
by up to 50\% below the free-electron value (Eq. (\ref{Eq_5})).

Just as Eq. (\ref{Eq_2}) {\em necessarily} enforces the
temperature scaling of all nonequilibrium thermal fluctuations,
so must Eq. (\ref{Eq_6}) enforce, in an
inhomogeneous mesoscopic contact,
the scaling of nonequilibrium fluctuations
with Coulomb suppression. Much more than that,
{\em Coulomb suppression is completely
determined by the equilibrium state}.
This has definite -- and observable -- physical consequences.

We have previewed some of the major, and completely generic,
results of the kinetic approach to mesoscopic transport.
In particular, we have highlighted the {\em microscopic}
structure of the fluctuations, and of their sum rules, as
being vital to the makeup of basic nonequilibrium processes. We now
discuss the technicalities of how this comes about.

\section*{3. Nonequilibrium Kinetics}

The focus of this section is on the conceptual structure
of the formalism, with mathematics in support.
First we recapitulate the open-system assumptions previewed
in Sec. 2. We link these to the essential {\em sum rules}
that the fluctuations of an electron gas must satisfy.
Then we show that
transmissive-diffusive models are in violation of at least one
of these constraints: the compressibility sum rule.
Finally, we survey our
rigorous kinetic solution for transport and noise.

Together with every other model of
current and noise in metals, including the
transmissive-diffusive description
\cite{imry,imry2,ferry,datta,djb,blbu},
our kinetic approach requires

\begin{itemize}
\item
an ideal thermal bath regulating the size of
energy exchanges with the conductor, while itself always remaining
in the equilibrium state;

\item
ideal macroscopic carrier reservoirs (leads)
in open contact with the conductor, without themselves
being driven out of their local equilibrium;

\item
absolute charge neutrality of the leads,
and overall neutrality of the intervening conductor.
\end{itemize}

\noindent
This standard scheme, consistently applied within the standard
framework of Boltzmann and, later, of Landau and Silin
\cite{pinoz,abri,bk},
puts specific and tight
constraints on the behavior of nonequilibrium current noise.

The electron gas in each asymptotic lead is unconditionally
neutral, and satisfies canonical identities for
{\em compressibility} and {\em perfect screening}
\cite{nozieres,pinoz}.
It has long been understood that they
embody the quantitative effects of degeneracy
(compressibility sum rule) and of Gauss' theorem
(perfect-screening sum rule).

Each criterion entails a precise
numerical relation between the mean charge {\em density}
and its {\em fluctuation}. Some feeling for the cardinal
role of the electron-gas sum rules, in noise and transport
together, can be gained by looking more closely at the
compressibility.

\subsection*{3.1 Compressibility: a Case Study in Sum Rules}

\subsubsection*{3.1.1 Compressibility and Electron-Gas Physics}

The compressibility sum rule links
the local physical density of the electron gas
$n(\varepsilon_{\rm F}({\bf r}))$ to the system's local,
screened polarization function
$\chi_0(q \ll k_{\rm F}, \omega = 0)$ in its adiabatic limit,
for wavelengths long relative to the inverse Fermi wavevector
$k_{\rm F}^{-1}$. Thus
\cite{pinoz}

\begin{equation}
\kappa
\equiv {1\over n^2} {{\partial n}\over {\partial \varepsilon_{\rm F}}}
= -{1\over n^2} \chi_0(0,0)
\equiv {2\over n^2 \Omega({\bf r})}
\sum_{\bf k} {{\Delta f^{\rm eq}_{\alpha}}\over k_{\rm B}T}.
\label{Eq_6.1}
\end{equation}

\noindent
Comparison with Eq. (\ref{Eq_2}) immediately shows the intimate
connection between this canonical equilibrium relation, and the
conservation of total fluctuation strength in a conductor taken
{\em out} of equilibrium.

Let us go to the global form of the compressibility rule,

\begin{eqnarray}
{\Omega \over N^2} {{\partial N}\over {\partial \mu}}
=&& {\Omega \over N^2k_{\rm B}T} \sum_{\bf r} \Omega({\bf r})
{\langle \Delta f^{\rm eq}({\bf r}) \rangle}
\cr
{\left. \right.} \cr
=&& {\Omega \over Nk_{\rm B}T}{{\Delta N}\over N},
\label{Eq_6.2}
\end{eqnarray}

\noindent
where the trace over spin and momentum states is
${\langle \cdot\cdot\cdot \rangle}
\equiv 2/\Omega({\bf r})\sum_{\bf k} \cdot\cdot\cdot$.
We make three observations.

\begin{itemize}
\item
In the limit of the classical gas, $\Delta N = N$.
Then the ideal-gas law shows that the right-hand
side is the inverse of the pressure. The pressure is the
thermodynamic bulk modulus, $\kappa^{-1}$.

In the quantum regime, $\Delta N < N$. The exclusion principle
keeps the electrons apart, making the system stiffer
so that (in a loose sense)
this is the Fermi-gas analog of van der Waals' hard-core model.

\item
Electron-hole symmetry is fundamental.
The microscopic basis of compressibility lies within
the same {\em electron-hole pair fluctuations}
that determine the structure of 
the polarization response function $\chi_0(q, \omega)$;
see also Footnote 4, Section 2.4 above.
The dynamical evolution of the electron-hole pair excitations
within $\chi_0(q, \omega)$ is {\em kinematically correlated}
by microscopic charge and current conservation
\footnote{
Current conservation is frequently discussed in the
sense of an augmented particle flux that includes the
{\em displacement} term associated with Poisson's
equation. The sum of the two has zero divergence;
consider the equation of conservation (continuity)
\\
\[
{\partial n\over \partial t} +
{\partial \over \partial {\bf r}}\cdot{\bf J} = 0,
\]
\\
\noindent
which comes from taking traces over ${\bf k}$ in the
equation of motion (refer to Eq. (\ref{9}) in the text).
Poisson's equation for the density gives
\\
\[
-4\pi e{\partial n\over \partial t}
= {\partial \over \partial t}
{\left(
{\partial \over \partial {\bf r}}\cdot \epsilon {\bf E}
\right)}.
\]
\\
\noindent
Then the continuity equation can be recast as
\\
\[
{\partial \over \partial {\bf r}}{~\cdot}{\left[
{\bf J} + {\partial \over \partial t}
{\left( -{\epsilon\over 4\pi e} {\bf E} \right)}
\right]}
= 0.
\]
\\
\noindent
The total flux is divergenceless {\em if and only if} the
originating equation of motion is gauge invariant.
There is no way to guarantee this result otherwise.
}
expressed through the electron-hole symmetry of transport
\cite{nozieres,pinoz};
refer also to Fig. 1(b).
The very same, inherently correlated, electron-hole
processes determine the noise
\cite{gdi,gdii}.

\item
For a nonuniform conductor, we must compute the total response
to a change in global chemical potential. As before
(recall Eq. (\ref{Eq_6c})) we now have

\begin{eqnarray}
{\Omega \over N^2} {{\delta N}\over {\delta \mu}}
=&&
{\Omega \over N^2k_{\rm B}T}
\cr
&&\times
\sum_{\bf r} \Omega({\bf r})
{ {\langle \Delta f^{\rm eq}({\bf r}) \rangle}\over
  {1 + {\displaystyle 
         { {\langle \Delta f^{\rm eq}({\bf r}) \rangle}\over k_{\rm B}T }
         { dU_0({\bf r})\over dn({\bf r}) }
       }
  }
}
\cr
{\left. \right.} \cr
\equiv&&
{\Omega \over Nk_{\rm B}T}{ {{\widetilde \Delta} N}\over N }.
\label{Eq_6.3}
\end{eqnarray}

\noindent
The internal Coulomb correlations, which
determine the local mean-field potential $U_0({\bf r})$, increase
the free energy of the electrons. This makes the electrons
stiffer yet, over and above the exchange correlations
evident in Eq. (\ref{Eq_6.2}).
It is a classic illustration of {\em Coulomb screening} at work,
and is obviously a major physical process in mesoscopic
structures whose spatial irregularities are large,
or else approach the scale of the
screening length
\cite{gdii}.
\end{itemize}

\subsubsection*{3.1.2 Compressibility and
Transmissive-Diffusive Phenomenology}

In Section 2 we discussed how the total fluctuation
strength of a mesoscopic conductor is invariant,
whether it is in equilibrium or not. We now see that this is
closely tied to the microscopics of the compressibility.
One should therefore ask for the corresponding behavior of
$\Delta N$ in a typical transmissive-diffusive model.

As a concrete example we take the noise theory of Martin and Landauer
\cite{marlan}
for an electronic conductor.
(We could as well have taken the Landauer-B\"uttiker
description
\cite{buett,blbu}.)
In addition, we recall that de Jong and Beenakker
have argued for an equivalence between the transmissive-diffusive
method and that of semiclassical (Boltzmann-Langevin)
theory
\cite{djb}.

The model of Ref.
\onlinecite{marlan}
builds up the current-current correlation function from the
set of all possible quantum-transmission events
through the conducting region.
We take their one-dimensional (1D) noise calculation
for a sample of length $L$ and (constant)
transmission probability ${\cal T}$.
Following their Eqs. (2.6)--(2.15),
the linear number fluctuation $\delta N$ can be computed.
\footnote{
The quantity $\delta N$ has nothing to do with
the electron-hole pair fluctuations intrinsic to the system.
It is generated purely by the carriers that enter and leave
the device, in Poissonian fashion.
As it turns out, $\delta N$ is simply proportional to the current
fluctuation for that model.
Note that this is characteristic of {\em all} models
built with the same transmissive-diffusive arguments
as Martin and Landauer's.
}
We arrive at the mean-square value

\end{multicols}

\begin{eqnarray}
\Delta N
\equiv&&
{\langle (\delta N)^2 - {\langle \delta N \rangle}^2 \rangle}
=
L{n\over 2\varepsilon_{\rm F}}
{\left[ {\cal T}^2 k_{\rm B}T +
{\cal T}(1 - {\cal T}){{\mu_{\rm S} - \mu_{\rm D}}\over 2}
{\rm coth}
{\left( {{\mu_{\rm S} - \mu_{\rm D}}\over 2k_{\rm B}T} \right)}
\right]}
\cr
{\left. \right.} \cr
=&& N{k_{\rm B}T\over 2\varepsilon_{\rm F}}
{\left[ {\cal T} + {{\cal T}(1 -{\cal T})\over 3}
{\left( {{\mu_{\rm S} - \mu_{\rm D}}\over 2k_{\rm B}T} \right)}^2
+ {\cal O}{\Bigl( ((\mu_{\rm S} - \mu_{\rm D})/2k_{\rm B}T)^4 \Bigr)}
\right]},
\label{Eq_7.0}
\end{eqnarray}

\begin{multicols}{2}

\noindent
where $n = 2k_{\rm F}/\pi$ is the 1D carrier density
while $\mu_{\rm S}$ and $\mu_{\rm D}$ are, respectively,
the ``chemical potentials'' assumed for the equilibrium state
of the source and drain leads.

Typically, as does {\em every other} diffusive model, the
Martin-Landauer theory supposes that the
EMF potential $eV$ fixes the difference between the source
and drain chemical potentials:

\begin{equation}
\mu_{\rm S} - \mu_{\rm D} \equiv eV.
\label{Eq_7.2}
\end{equation}

\noindent
It follows directly that, to leading order in the EMF, the
diffusively driven Martin-Landauer theory predicts

\begin{equation}
{\Delta N\over N}
= {\left[ {\Delta N\over N} \right]}^{\rm eq}
{\left[ {\cal T} + {{\cal T}(1 - {\cal T})\over 3}
{\left( {eV\over 2k_{\rm B}T} \right)}^2 \right]},
\label{Eq_7.1}
\end{equation}

\noindent
in which

\[
{\left[ {\Delta N\over N} \right]}^{\rm eq}
= {k_{\rm B}T\over 2\varepsilon_{\rm F}}
\]

\noindent
is the 1D equilibrium ratio of the
total mean-square number fluctuation to total carrier number.

If Eqs. (\ref{Eq_1}) and (\ref{Eq_2}) are correct,
as we will prove, then Eq. (\ref{Eq_7.1})
violates the compressibility sum rule. Therefore the
fluctuation structure of this diffusive model
also violates number conservation.

This is the cost of neglecting electron-hole symmetry in
the construction of pseudodiffusive transport.
All transmissive-diffusive models do this without exception.
For an interesting comment
on such violations, see Ref.
\onlinecite{blbu},
Eq. (51) and subsequent paragraph.

One can now answer the two core
questions posed in our Introduction:

\begin{itemize}
\item
{\bf Q}. Do transmissive-diffusive theories fully
respect {\em all} of the essential physics of the electron gas?
\\
{\bf A}. No.

\item
{\bf Q}. If not, {\em why} not?
\\
{\bf A}. There are two reasons.

(i) The total fluctuation $\Delta N$ in the transmissive-diffusive
models depends on the transport
parameter ${\cal T}$. It vanishes with ${\cal T}$.
As we have seen, the compressibility is an
equilibrium property insensitive to external sources
of elastic scattering (such as potential barriers) which fix ${\cal T}$.
Thus Eq. (\ref{Eq_7.1}) cannot recover the physical compressibility,
{\em even in the elementary zero-field limit of such models}.
Nor is it possible to invoke Coulomb suppression to account for
the spurious dependence on ${\cal T}$. This unphysical
result is for a uniform, free-electron model.

\noindent
(ii) Such theories grossly mistreat the role of the equilibrium
state in each bounding reservoir. The relevant thermodynamic
chemical potentials are not at all $\mu_{\rm S}$ and
$\mu_{\rm D}$, but the {\em undisturbed} equilibrium values.
These remain {\em locally invariant} within each lead.
Only then can the electron reservoirs fulfill their role: to
stabilize, screen, and confine the nonequilibrium fields and
their fluctuations within the active region
\cite{fenton,wims,gdi,kk}.
(At zero current, of course, each lead chemical potential
aligns with the global $\mu$.)

In view of the prevalence of pseudodiffusive thinking,
one cannot reassert sufficiently strongly the overwhelming
physical importance of this unconditional constraint:
the reservoirs' chemical potentials are {\em always local}
and {\em always undisturbed}.

Unequivocally, these {\em local-equilibrium} quantities
are the only ones that can appear in the transport description.
That is the only rule compatible with the microscopic
structure of the electron gas, both in the sample and
its stabilizing leads.

\end{itemize}

\subsection*{3.2 Nonequilibrium Carrier Distribution}

To confirm the fluctuation sum rules Eqs. (\ref{Eq_2}) and
(\ref{Eq_6c}), disconfirming in the process
the counterfeit fluctuation equation (\ref{Eq_7.0}),
we must show that the nonequilibrium
carrier fluctuations are {\em linear functionals} of the
equilibrium ones. From this follow all of the results that
we have already discussed.

We will need the one-electron equilibrium distribution. It is

\begin{equation}
f^{\rm eq}_{\alpha} =
{\left[
1 + \exp{\left(
{{\varepsilon_{\bf k} + U_0({\bf r}) - \mu}
\over k_BT}
\right)}
\right]}^{-1}.
\label{7}
\end{equation}

\noindent
The conduction-band energy $\varepsilon_{\bf k}$
can vary (implicitly) with ${\bf r}$ if the local band
structure varies, as in a heterojunction.
The mean-field potential $U_0({\bf r})$ vanishes asymptotically
in the leads, and
satisfies the self-consistent Poisson equation
($\epsilon$ is the background-lattice dielectric constant)

\begin{equation}
\nabla^2 U_0 \equiv
e{\partial\over {\partial {\bf r}}} \cdot {\bf E}_0
= -{{4 \pi e^2}\over \epsilon}
{\Bigl( \langle f^{\rm eq}({\bf r}) \rangle - n^+({\bf r}) \Bigr)}
\label{8}
\end{equation}

\noindent
in which, for later use,  ${\bf E}_0({\bf r})$
is the internal field in equilibrium
(recall that a nonuniform system sustains nonzero internal fields).
The (nonuniform) neutralizing
background density $n^+({\bf r})$ goes to the same constant
value, $n$, as the electrons in the (uniform) leads.

We study the semiclassical Boltzmann--Landau-Silin equation.
There is a substantial body of work, at every level, on this
transport equation. Among the analyses that we have found
most useful, we cite Refs.
\onlinecite{nvk,ks,ggk}
for Boltzmann-oriented kinetic descriptions and Refs.
\onlinecite{pinoz,bk,sv}
for more Fermi-liquid-oriented ones in the spirit of
Landau and Silin.

The kinetic equation, subject to
the total internal field ${\bf E}({\bf r}, t)$, can be written as

\begin{equation}
{\left( { \partial\over {\partial t} }
+ D_{\alpha}[{\bf E}({\bf r}, t)]
\right)} f_{\alpha}(t)
= -{\cal W}_{\alpha}[f].
\label{9}
\end{equation}

\noindent
Here $D_{\alpha}[{\bf E}] \equiv
{\bf v}_{\bf k}{\cdot}{ \partial/{\partial {\bf r}} }
- { (e{\bf E}/\hbar})
{\cdot}{ \partial/{\partial {\bf k}} }$ is the convective operator
and ${\cal W}_{\alpha}[f]$ is the collision operator, whose kernel
(local in real space) is assumed to
satisfy detailed balance, as usual
\cite{nvk}.
Even for single-particle impurity
scattering, Pauli blocking of the outgoing scattering states still
means that ${\cal W}$ is generally nonlinear in the
nonequilibrium function $f(t)$.

Since we follow the standard Boltzmann--Landau-Silin formalism
\cite{pinoz,nvk,ggk},
all of our results will comply with the conservation laws.
The nonlinear properties of these results extend as far as the
inbuilt limits of the semiclassical framework. These go
much further than any model restricted to the weak-field domain.
Since we rely expressly on the whole fluctuation
structure provided by Fermi-liquid theory
\cite{pinoz}, all of the fundamental sum rules are incorporated.

We develop our theory for
the steady-state distribution $f_{\alpha}$
out of equilibrium by expressing it
as an explicit functional of the equilibrium distribution.
The latter satisfies

\begin{equation}
D_{\alpha}[{\bf E}_0({\bf r})]
f^{\rm eq}_{\alpha}
= 0
= -{\cal W}_{\alpha}[f^{\rm eq}],
\label{10}
\end{equation}

\noindent
the second equality following by detailed balance.
Subtract the corresponding sides of Eq. (\ref{10})
from both sides of the time-independent
version of Eq. (\ref{9}). On introducing the difference
function
$g_{\alpha} \equiv
f_{\alpha} - f^{\rm eq}_{\alpha}$,
one obtains


\begin{eqnarray}
\sum_{\beta}{}&&{}
{\Bigl( {\cal I}_{\alpha \beta}
D_{\beta}[{\bf E}({\bf r}_{\beta})]
+ {\cal W}'_{\alpha \beta}[f] \Bigr)}
g_{\beta}
\cr
=
&&
 { {e[{\bf E}({\bf r}) - {\bf E}_0({\bf r})]}\over \hbar } \cdot
{ {\partial f^{\rm eq}_{\alpha}}\over {\partial {\bf k}} }
- {\cal W}''_{\alpha}[g].
\label{11}
\end{eqnarray}


\noindent
The unit operator in Eq. (\ref{11}) is

\[
{\cal I}_{\alpha \alpha'} \equiv
{\left[ {\delta_{\bf kk'}\over \Omega({\bf r})} \right]}
{\left[ \Omega({\bf r}) \delta_{\bf rr'} \right]}
\]

\noindent
and the linearized operator ${\cal W}'[f]$ is
the variational derivative

\[
{\cal W}'_{\alpha \alpha'}[f] \equiv
{ {\delta {\cal W}_{\alpha}[f]}\over {\delta f_{\alpha'}} }.
\]

\noindent
Last, the collision term

\[{\cal W}''_{\alpha}[g] \equiv
{\cal W}_{\alpha}[f] - {\cal W}_{\alpha}[f^{\rm eq}]
- \sum_{\beta}{\cal W}'_{\alpha \beta}[f] g_{\beta}
\]

\noindent
carries the residual nonlinear contributions.
Although ${\cal W}_{\alpha}[f^{\rm eq}]$ is identically zero by
detailed balance, ${\cal W}'_{\alpha \alpha'}[f^{\rm eq}]$ is not.
We must formally keep the equilibrium quantity,
via ${\cal W}''_{\alpha}[g]$, on the right-hand
side of Eq. (\ref{11}) because we will require its variational
derivative.

Global neutrality enforces the fundamental constraint

\begin{equation}
\sum_{\alpha} g_{\alpha}
= \sum_{\bf r} \Omega({\bf r}) {\langle g({\bf r}) \rangle} = 0.
\label{11.0}
\end{equation}

\noindent
We need not elaborate; Eq. (\ref{11.0}) is the
immediate consequence of the general boundary conditions
introduced at the Section's beginning. From it, all of
the sum-rule results are derived.

The leading right-hand term in Eq. (\ref{11}) is responsible
for the functional dependence of $g$ on the equilibrium
distribution (this is important because dependence
on {\em equilibrium-state} properties
carries through to the variationally derived
steady-state fluctuations).
The electric-field factor can be written as

\[
{\bf E}({\bf r}) - {\bf E}_0({\bf r})
\equiv {\widetilde {\bf E}}({\bf r})
= {\bf E}_{\rm ext}({\bf r}) + {\bf E}_{\rm ind}({\bf r}),
\]

\noindent
where ${\bf E}_{\rm ext}({\bf r})$ is the external
driving field,
and the induced field ${\bf E}_{\rm ind}({\bf r})$
obeys

\begin{equation}
{\partial\over {\partial {\bf r}}}{\cdot} {\bf E}_{\rm ind}
= -{{4\pi e}\over \epsilon} {\langle g({\bf r}) \rangle}.
\label{12}
\end{equation}

\noindent
Equation (\ref{12}) guarantees that $g$ vanishes in the
equilibrium limit. This maintains the so-called adiabatic
connection of the nonequilibrium solution $f$ to $f^{\rm eq}$.

\subsection*{3.3 Nonequilibrium Fluctuations; Analytical Form}

Now we consider the nonequilibrium fluctuation
$\Delta f_{\alpha}(t)$. It satisfies the
(well documented) linearized equation of motion
\cite{ks,ggk}

\begin{equation}
\sum_{\beta}
{\left[
{\cal I}_{\alpha \beta}
{\left( {\partial\over {\partial t}} +
D_{\beta}[{\bf E}({\bf r}_{\beta})] \right)}
+ {\cal W}'_{\alpha \beta}[f]
\right]} \Delta f_{\beta}(t)
= 0.
\label{13}
\end{equation}

\noindent
This equation remains subject to the same unconditional
boundary constraints that we have discussed.
In the Landau Fermi-liquid regime,
it generates all of the dynamical
properties of the fluctuating electron gas.
Once it is solved, {\em all} of the physical properties
of the current fluctuations can be computed.

For the adiabatic $t \to \infty$ limit,
$\Delta f_{\alpha}(t) \to \Delta f_{\alpha}$
represents the average strength of
the spontaneous background fluctuations,
induced in the steady state by the ideal thermal bath.
It is one of two essential
components that determine the dynamical fluctuations.
The other component is the dynamical
Green function for the
inhomogeneous version of Eq. (\ref{13}).
See Ref.
\onlinecite{gdii}.

In a strongly degenerate system $\Delta f_{\alpha}$
dictates the explicit $T$-scaling of all thermally
based noise 
through its functional dependence on
the equilibrium distribution
$\Delta f^{\rm eq}_{\bf k}({\bf r})$.
We saw this in Eqs. (\ref{Eq_3}) and (\ref{Eq_6}).
Now we prove it.

Define the variational derivative

\begin{equation}
G_{\alpha \alpha'}[f] \equiv
{\Biggl.
{ {\delta g_{\alpha}}\over {\delta f^{\rm eq}_{\alpha'}} }
\Biggr|}_{\bf E}.
\label{11.1}
\end{equation}

\noindent
This operator
obeys a steady-state equation obtained from Eq. (\ref{11})
by taking variations on both sides.
Note that we restrict the variation by keeping
the total internal field constant. This provides us with
the nonequilibrium Fermi-liquid response of the system
(dominated by degeneracy). The self-consistent Coulomb
field fluctuations can be obtained, systematically,
by lifting the variational restriction. See our Ref.
\cite{gdii}.

The equation for $G$ is

\end{multicols}

\begin{equation}
\sum_{\beta}
{\Bigl( {\cal I}_{\alpha \beta}
{\cal D}_{\beta}[{\bf E}({\bf r}_{\beta})]
+ {\cal W}'_{\alpha \beta}[f] \Bigr)}
G_{\beta \alpha'}
= {\cal I}_{\alpha \alpha'}
{ e{\widetilde {\bf E}}({\bf r'})\over \hbar } \cdot
  {\partial\over {\partial {\bf k'}}}
- {\cal W}'_{\alpha \alpha'}[f]
+ {\cal W}'_{\alpha \alpha'}[f^{\em eq}].
\label{11.2}
\end{equation}

\begin{multicols}{2}

\noindent
The explicit and  closed form for $G$, which we do not give here,
is obtained from knowledge the dynamical Green function
for the linearized equation of motion, Eq. (\ref{13})
\cite{gdi,gdii}.
The main point, of utmost physical importance, is that the expression

\begin{equation}
\Delta f_{\alpha} =  \Delta f^{\rm eq}_{\alpha}
+ \sum_{\alpha'} G_{\alpha \alpha'} \Delta f^{\rm eq}_{\alpha'}
\label{14}
\end{equation}

\noindent
satisfies the steady-state form of Eq. (\ref{13}) {\em exactly}.
In the form above, $\Delta f_{\alpha}$
is the definitive solution for the steady-state,
mean-square fluctuation in nonequilibrium transport.

Eqs. (\ref{Eq_2}) and (\ref{Eq_3}) can now be confirmed in steady state
by invoking the unconditional neutrality
of $g$; see Eq. (\ref{11.0}). This immediately implies

\begin{mathletters}
\label{14.0}

\begin{equation}
\sum_{\alpha} G_{\alpha \alpha'} = 0 {~~~}{\rm for ~all}{~}\alpha'.
\label{14.0a}
\end{equation}

\noindent
Hence

\begin{eqnarray}
\sum_{\alpha} \Delta f_{\alpha}
=&&
\sum_{\alpha} \Delta f^{\rm eq}_{\alpha}
+ \sum_{\alpha'} {\left( \sum_{\alpha} G_{\alpha \alpha'} \right)}
\Delta f^{\rm eq}_{\alpha'}
\cr
{\left. \right.} \cr
=&&
\sum_{\alpha} \Delta f^{\rm eq}_{\alpha},
\label{14.0b}
\end{eqnarray}

\end{mathletters}

\noindent
which establishes the static form of the compressibility sum rule;
an {\em exact} constraint on the nonequilibrium carrier fluctuations
in a mesoscopic conductor. It holds under very general
boundary conditions and modes of scattering (quasiparticle
interactions are included in the collision integral ${\cal W}[f]$,
as well as external collision processes). A well controlled
theory of mesoscopic noise {\em must take the compressibility
sum rule into account} at the very least (there are several others
\cite{pinoz}).

The stationary fluctuation properties of a driven system are
intimately connected to its dynamic response. We end this
technical discussion with a description of the noise
spectral density.

\subsection*{3.4 Nonequilibrium Fluctuations: Dynamics}

\subsubsection*{3.4.1 Dynamic Fluctuation Structure}

The time-dependent Green function is the variational derivative
(with Coulomb effects restricted)

\begin{equation}
R_{\alpha \alpha'}(t - t')
\equiv \theta(t - t')
{\Biggl.
{{\delta f_{\alpha}(t)}\over {\delta f_{\alpha}(t')}}
\Biggr|}_{\bf E};
\label{15}
\end{equation}

\noindent
$\theta(t - t')$ is the Heaviside unit-step function.
In the low-field limit, the Fourier transform of $R$
is closely related to the internal makeup
of the dynamic polarization $\chi_0(q,\omega)$
\cite{pinoz}.
It can be solved routinely
\cite{sw1,sw2,ks,ggk}.

The exact solution to the equation of motion
for the dynamical fluctuation, Eq. (\ref{13}), is

\begin{mathletters}
\label{15.0}

\begin{equation}
\Delta f_{\alpha}(t)
= \sum_{\alpha'} R_{\alpha \alpha'}(t) \Delta f_{\alpha'}.
\label{15.0a}
\end{equation}

\noindent
The conserving nature of $R$ implies that
$\sum_{\alpha} R_{\alpha \alpha'}(t) = 1$ for all $\alpha'$.
It follows that
\cite{gdi}

\begin{equation}
\sum_{\alpha} \Delta f_{\alpha}(t)
= \sum_{\alpha'} \sum_{\alpha} R_{\alpha \alpha'}(t) \Delta f_{\alpha'}
= \sum_{\alpha'} \Delta f_{\alpha'}.
\label{15.0b}
\end{equation}

\end{mathletters}

\noindent
With Eq. (\ref{15.0b}) and Eq. (\ref{14.0b}) in association,
we complete the promised derivation of Eq. (\ref{Eq_2}), which
essentially fixes the dynamic global compressibility
in a mesoscopic conductor out of equilibrium.

From the point of view of microscopic analysis, our derivation
is {\em entirely standard} and thus definitive. The only way to
circumvent its negative implication for transmissive-diffusive
theory, would be to show that its long-established basis in
electron-gas physics -- going back almost a century -- is erroneous.

The proof for the exact Coulomb-suppressed compressibility
Eq. (\ref{Eq_6.3}) develops along parallel lines, apart from
the added self-consistency feature. It is fully set out
in Ref.
\onlinecite{gdii}. 

\subsubsection*{3.4.2 Current-Current Correlation}

For the current autocorrelation
we require the {\em transient} part of the propagator $R$
\cite{ks,ggk},

\begin{equation}
C_{\alpha \alpha'}(t) = R_{\alpha \alpha'}(t)
- R_{\alpha \alpha'}(t \to \infty).
\label{15.1}
\end{equation}

\noindent
The transient propagator carries all the dynamical correlations.
As is standard practice
\cite{ks,ggk},
the flux autocorrelation can be written down directly
in terms of $C$ and $\Delta f$:


\begin{eqnarray}
S_{JJ}({\bf r}, {\bf r'}; t)
\equiv
&&
{2\over \Omega({\bf r}) \Omega({\bf r'})}
\sum_{\bf k} \sum_{\bf k'}
[-e(v_x)_{\bf k}] C_{\alpha \alpha'}(t)
\cr
&&{~~~ ~~~ ~~~ ~~~ ~~~ ~~~ ~~~}\times
[-e(v_x)_{\bf k'}] \Delta f_{\alpha'},
\label{15.2}
\end{eqnarray}


\noindent
where for illustration we select the $x$-components of the velocities.
(This is the most relevant term for a uniform conductor with
the driving field acting along the $x$-axis.)

Let us outline the physical meaning of Eq. (\ref{15.2}). In steady
state, the average fluctuation strength is $\Delta f$. Once
a spontaneous thermal fluctuation (with this strength)
arises within the system, it evolves and decays
as a result of collisional processes.
The transient evolution, and its characteristic time constant,
are given by $C$. There are three parts to the exercise:

\smallskip
(a) the object ${\bf v'}\Delta f'$ represents,
in the mean, a spontaneous flux fluctuation.

(b) After time $t$, the fluctuation has evolved to
$C(t){\bf v'}\Delta f'$.

(c) The velocity autocorrelation that describes
this dynamical process is ${\bf v}C(t){\bf v'}\Delta f'$.

\subsubsection*{3.4.3 Temperature Scaling}

Since $S_{JJ}$ scales with $\Delta f$, which itself scales with
$\Delta f^{\rm eq}$, our conclusion for the current-current
fluctuation in a degenerate conductor is inescapable.

\begin{itemize}
\item
{\em In a metallic system, the current-current correlator
always scales with temperature $T$}.
\end{itemize}

\noindent
This strict result
leaves transmissive-diffusive models
\cite{blbu}
in a difficult, indeed untenable, position.
On the one hand, their current-current correlator
must revert to the mandatory Johnson-Nyquist form at
low fields. This is canonically proportional to $T$. On the other
hand, consider the high-field, low-frequency limit of the noise
spectral density in the theory of Ref.
\cite{marlan},
whose form is identical for all of the theories in question:


\begin{eqnarray}
{\cal S}({}&&{}V; \omega\!=\!0)
\cr
{\left. \right.} \cr
=&&
4{e^2{\cal T}\over {\pi \hbar}}
{\left[ {\cal T}k_{\rm B}T +
(1\!-\!{\cal T}){{\mu_{\rm S}\!-\!\mu_{\rm D}}\over 2}
{\rm coth}
{\left( {{\mu_{\rm S}\!-\!\mu_{\rm D}}\over 2k_{\rm B}T} \right)}
\right]}
\cr
{\left. \right.} \cr
\to&& 4{e^2{\cal T}\over {\pi \hbar}}
{\left( {\cal T}k_{\rm B}T
+ (1 - {\cal T}){eV\over 2}
\right)}.
\label{16}
\end{eqnarray}


\noindent
The dominant term is the last one on the right-hand side,
ascribed to shot-noise processes. It does not
scale with temperature, as required by the compressibility
sum rule.
It follows that the current-current
correlator in such a model, on which the derivation of
${\cal S}(V,\omega)$ is based,
cannot be the canonical one, Eq. (\ref{15.2})
\cite{ks,ggk}.
Hence

\begin{itemize}
\item
{\em Equation (\ref{16}) and the 
Landauer-B\"uttiker-Imry phenomenology
that leads directly to it, are in manifest and
irreconcilable conflict with canonical microscopics.}
\end{itemize}

Does the strict $T$-scaling of $S_{JJ}$
mean that shot noise is an ill-defined concept in
the kinetic description of a degenerate mesoscopic conductor?
Not at all. Shot noise is a real effect

The canonically obtained form
for $S_{JJ}$ -- with its $T$-scaling -- clearly
implies that shot-noise
fluctuations of a degenerate conductor
must have a physical origin, and behavior,
entirely distinct from its thermal fluctuations.
Therefore

\begin{itemize}
\item
{\em Shot noise must have a microscopic description
entirely distinct from that for ``hot-electron'' noise,
incorporated within Eq. (\ref{15.2}).}
\end{itemize}

\noindent
We do not give the kinetic-theoretical treatment of shot noise
in the present review. Such a treatment is available in our Refs.
\onlinecite{ajp}
and
\onlinecite{9809339}.
In essence, shot noise is a time-of-flight process
measured between the device boundaries.
(Its intuitive meaning is well depicted by Martin and Landauer
\cite{marlan},
though in a formalism incompatible with the electron gas.)

Shot noise involves
discrete changes in the total carrier number $N$.
By contrast, thermal noise is a volume-distributed process.
It involves continuous changes of internal energy.
The two are numerically very different, though both share
the same variational, microscopic building blocks:
$C$ and $\partial f/\partial \mu$
or, in the case of shot noise, $\partial f/\partial N$.

\subsection*{3.5 Coda}

Our primary goal is met. We have described the
structure and physical consequences of a kinetic approach to
noise that is strictly conserving. The intent of our first-principles
mesoscopics program is aptly put by Imry and Landauer
\cite{imry}:

\smallskip
\noindent
{\em Kubo's linear-response theory is essentially an extended theory
of polarizability. Some supplementary hand-waving is needed to
calculate a dissipative effect such as conductance, for a sample
with boundaries where electrons enter and leave... After all, no
theory that ignores the interfaces of a sample to the rest of its
circuit can possibly calculate the resistance of such a sample of
limited extent.}
\smallskip

\noindent
No more need be said, save for four incidental remarks.

\begin{itemize}
\item
A properly constituted conductance and fluctuation theory
of the electron gas {\bf IS} a theory of the polarizability
\cite{gdi,kk}.
A polarization-based model is not a matter of taste;
the physics of electron-hole processes in the electron gas
\cite{pinoz}
demands it. All self-styled alternatives are nonconservative.
Furthermore, the Kubo conductance formula
\cite{kubo}
emerges directly from an axiomatic derivation of the
fluctuation-{\em dissipation} theorem
(an accomplishment beyond Ref.
\onlinecite{imry}
and its like).

\item
No hand-waving, supplementary {\em or otherwise},
is needed to calculate dissipation. That is automatic
for a model (such as Kubo's) which guarantees its
fluctuation-dissipation theorem from first principles
\cite{kk},
rather than having to take it on faith.

\item
It is not merely well known how to include dissipation;
it is {\em obligatory} to do so explicitly, microscopically,
and in perfect harmony with gauge invariance.
Even the humble Drude model -- with its supposedly
``primitive'' understanding -- easily achieves that much, at least
\cite{fenton,kk,quinn,mermin}.
The same cannot be said of purely intuitive schemes.

\item
Transmissive-diffusive phenomenology {\em itself}
ignores the avowedly crucial interface physics. That is
why it mistreats the canonical compressibility so grossly.
\end{itemize}

Kinetic theory, unlike the pseudodiffusive mindset,
respects the sum rules that have been established
-- universally and decades ago
\cite{nozieres,pinoz,abri} --
as definitive expressions of the Fermi-liquid origin of electron-hole
correlations. They govern two phenomena, {\em conduction} and {\em noise}.
It remains to give a major application of what is,
in every way, a thoroughly conventional microscopic approach:
the behavior of high-current thermal noise in one-dimensional
ballistic wires and quantum point contacts.

\section*{4. Ballistic Noise}

We review our results for 1D ballistic noise,
reported recently and more fully in Ref.
\onlinecite{fnl}.
That work has the complete details.
The quantity that we wish to calculate is the
long-time limit of the thermal-noise correlation

\begin{equation}
{\cal S}(V) \equiv 4\int^{\infty}_0 dt
\int^{L/2}_{-L/2} dx \int^{L/2}_{-L/2} dx'
{~}{S_{JJ}(x, x'; t)\over L^2}
\label{30}
\end{equation}

\noindent
for a 1D mesoscopic conductor of length $L$. Our calculation
covers both diffusive and ballistic cases, but we focus on the latter.

\subsection*{4.1 Transport Problem}

Recall Fig. 2 for a mesoscopic wire in close electrical
contact with its reservoirs. The wire is {\em uniform}, except
possibly in the restricted fringing regions where the current,
as it is injected and extracted, strongly
perturbs the local electrons. This induces a net
charge displacement, responsible for Landauer's resistivity dipole
\cite{ldr57},
which is also the EMF.
Under the conditions of strong screening and
{\em phase breaking} imposed by the reservoirs, it can be
argued that the carriers crossing the active region have
no detailed memory of the boundary disturbances. Within
the wire, they are Markovian and obey
the spatially homogeneous form of the kinetic equation,
Eq. (\ref{9}).

Furthermore, the explicit
presence of the current source and sink
\cite{sols}, with their
associated regions of strong relaxation by scattering, means physically
that the dissipative effects of {\em inelastic collisions}
must be explicitly represented. Once again, we stress
that vague appeals to dissipative relaxation in the
leads' asymptotic equilibrium state
\cite{imry}
avail nothing to the description of
real {\em driven} mesoscopic transport.

The ballistic kinetic equation is


\begin{eqnarray}
{\partial f_k\over \partial t}
+ {eE\over \hbar} {\partial f_k\over \partial k}
=&&
-{1\over \tau_{\rm in}(\varepsilon_k)}
{\left( f_k(t) -
{{\langle \tau_{\rm in}^{-1} f(t) \rangle}\over 
 {\langle \tau_{\rm in}^{-1} f^{\rm eq} \rangle}}
f^{\rm eq}_k
\right)}
\cr
{\left. \right.} \cr
&&
-{1\over \tau_{\rm el}(\varepsilon_k)}
{ {f_k(t) - f_{-k}(t)}\over 2 }.
\label{31}
\end{eqnarray}


\noindent
The uniform driving field is $E = V/L$. For the collision
operator we adopt a Boltzmann-Drude form that
includes the inelastic collision
time $\tau_{\rm in}(\varepsilon_k)$ as well as the elastic time
$\tau_{\rm el}(\varepsilon_k)$. The structure of the
inelastic collision contribution on the right-hand side
 automatically ensures charge and current conservation.

The solution to Eq. (\ref{31}) can be written down analytically
for collision times that are independent of particle energy
\cite{fnl}. In the sense of our open-system kinetics, the
1D wire is collision-free (that is, ballistic) when
the dominant mean free paths $v_{\rm F}\tau_{\rm in}$
and $v_{\rm F}\tau_{\rm el}$ (for Fermi velocity $v_{\rm F}$)
are at their {\em maximum span}. That happens only
when both are equal to the ``ballistic length'' $L$ between the
regions of strong relaxation, at the current entry and exit points.
The ballistic length is therefore set by the {\em longest}
mean free path in the problem, which cannot be greater than
the distance between the sites for relaxation.

This ballistic condition leads straight to Landauer's
ideal quantized conductance
\cite{fnl}:

\begin{equation}
G = {I\over V} = {e^2\over \pi \hbar}
\label{32}
\end{equation}

\noindent
for a single, occupied subband within the (open) 1D wire.
When conditions are nonideal, so that the wire is 
either ``elastic--diffusive''
($\tau_{\rm el} < \tau_{\rm in} = L/v_{\rm F}$)
or ``inelastic--dissipative''
($\tau_{\rm in} < \tau_{\rm el} = L/v_{\rm F}$),
then

\begin{equation}
G = {e^2\over \pi \hbar}
{\left(
1 - {|\tau_{\rm in} - \tau_{\rm el}|\over {\tau_{\rm in} + \tau_{\rm el}}}
\right)}.
\label{33}
\end{equation}

\noindent
The second ratio on the right-hand side plays the role of
the Landauer-B\"uttiker transmission probability ${\cal T}$
except that inelastic effects are fully included; the
transmissive-diffusive treatment of ${\cal T}$ admits only
coherent, purely elastic, scattering
\cite{imry,blbu}.

\subsection*{4.2 Ballistic Hot-electron Noise}

Nonideality in the 1D conductance is well documented in many
ballistic tests of Landauer's quantized formula.
Nonideal conductance appears even in the most refined state-of-the-art
measurements, notably the recent ones by de Picciotto {\em et al}.
\cite{depic}.
It is of great interest to predict the corresponding
nonideal behavior of the nonequilibrium thermal noise.

Our conserving kinetic theory, worked out according to the
methods described in Sec. 3, results in a noise spectral density
that is exact for the transport model of Eq. (\ref{31}).
Expressed as the thermal hot-electron excess noise within a
given subband of carrier states in the 1D conductor,
say the $i$th one, it is


\end{multicols}

\begin{eqnarray}
{\cal S}^{\rm xs}_i(V)
=
&& 
{\cal S}_i(V) - 4G_ik_{\rm B}T
=
{\kappa_i\over \kappa^{\rm cl}_i}
{2e^2I^2\over G_i m^* L^2}
{\left(
\tau_{{\rm in};i}^2
+ 2{ {\tau_{{\rm el};i} \tau_{{\rm in};i}^2}\over
     {\tau_{{\rm in};i} +\tau_{{\rm el};i}} } 
-  { {\tau_{{\rm el};i}^2 \tau_{{\rm in};i}^2}\over
     (\tau_{{\rm in};i} +\tau_{{\rm el};i})^2 }
\right)},
\label{34}
\end{eqnarray}

\begin{multicols}{2}

\noindent
where $\kappa^{\rm cl}_i = 1/n_i k_{\rm B}T$ is
the classical compressibility. The subscripts ``$i$''
on all quantities identify the subband;
for instance (in the case that inelastic phonon emission
modifies the ideal conductance), we
have

\[
G_i = {e^2\over \pi \hbar}{2\tau_{{\rm in};i}\over
(\tau_{{\rm in};i} + \tau_{{\rm el};i})}
\].

Note once again the overall $T$-{\em scaling} of the excess noise
in Eq. (\ref{34}). This is due to its obviously
intimate link with the compressibility, entering via the factor
$\kappa_i/\kappa^{\rm cl}_i = \Delta n_i/n_i$.
It is the necessary consequence of microscopic conservation.
As we saw above, transmissive-diffusive approaches
are seriously defective in that essential regard. 

We make several comments on the nature of the
ballistic hot-electron spectral density.

\begin{itemize}

\item
The dependence on collision times (the last right-hand factor in
Eq. (\ref{34})) is greatly enhanced over that of $G_i$.
As the 1D structure is taken beyond its low-current regime,
the excess thermal noise should reflect much more strongly
the onset of nonideal behavior.

\item
The nonlinear form of ${\cal S}_i^{\rm xs}(V)$ as
a function of V shows that it is not shot noise.
This is not too astonishing, in view of our earlier discussion.

\item
When inelastic effects are dominant, $\tau_{{\rm in};i}$
is small and makes the ratio
${\cal S}_i^{\rm xs}/G_i$ small. Conversely, when
$\tau_{{\rm in};i}$ becomes artificially large
(the inelastic mean free path is made
to exceed its maximum physical limit, $L$),
then ${\cal S}_i^{\rm xs}/G_i$ diverges.

This divergence indicates that noise models relying on
elastic scattering {\em alone}, for their current-voltage response,
are thermodynamically unstable
beyond the zero-field limit. There is simply no mechanism
for field-excited carriers to shed excess energy.
The excess then manifests as an uncontrolled broadening of
their distribution, and a very large thermal noise spectrum. 

\item
In the highly degenerate regime, the noise spectrum scales
as $\kappa_i/\kappa^{\rm cl}_i = k_{\rm B}T/2(\mu - \varepsilon_i)$,
where $\varepsilon_i$ is the subband threshold energy.
For a well filled subband, the noise is strongly suppressed.
In the classical limit, ${\cal S}_i^{\rm xs}$
becomes {\em independent} of temperature as
$\kappa_i/\kappa^{\rm cl}_i \to 1$.

\end{itemize}

Experiments on 1D ballistic wires or on quantum point contacts
are designed so that the subband occupancies in their structures
can be systematically changed via a gate-control potential
\cite{rez,depic}.
We have described the marked behavioral change in the
hot-electron noise as a function of subband density $n_i$.
This suggests some intriguing possibilities for excess-noise
measurements in 1D wires, particularly at higher
source-drain fields.

\subsection*{4.3 Results}

The following scenario now unfolds.
When a subband is depopulated
(classical limit; $\mu - \varepsilon_i \ll k_{\rm B}T$),
the factor $\kappa_i/\kappa^{\rm cl}_i$ of ${\cal S}^{\rm xs}_i$
is at its maximum value, unity.
At the same time, the conductance $G_i$ is negligible, since it
scales with $n_i$ which vanishes. The vanishing of $G_i$
means that there is little spectral strength in the noise.

As we cross the subband threshold (with $G_i$ now rising from
nearly zero up to $e^2/\pi \hbar$), the factor
$\kappa_i/\kappa^{\rm cl}_i$ starts to drop
in magnitude. Well above the threshold
(quantum limit; $\mu - \varepsilon_i \gg k_{\rm B}T$),
$G_i$ is a maximum, but
$\kappa_i/\kappa^{\rm cl}_i
= k_{\rm B}T/2(\mu - \varepsilon_i) \ll 1$.
Again there is little spectral strength.

We see that ${\cal S}_i^{\rm xs}$ must pass
through a maximum close to the energy threshold 
$\mu = \varepsilon_i$. Below it, the noise is that of
a low-density gas of classical carriers. Above, it is
that of a highly degenerate Fermi system.

Our results are shown in Fig. 5 for a 1D wire
with two subbands
\onlinecite{fnl}.
The peaks in the hot-electron noise are dramatic,
somewhat unexpected, and much less likely to be resolved
in two- or three-dimensional systems. The peak structures
are due directly to the strong influence of
electron degeneracy (indeed, of the compressibility sum rule)
in 1D metallic systems.

In the same Figure, we display the corresponding
ideal-noise spectral density of transmissive-diffusive theory
\cite{blbu}
(refer to Eq. (\ref{16}) in the previous Section).
As we have shown, that approach badly violates
the compressiblity sum rule and hence charge conservation.
In any case, at high fields it is overshadowed by the
hot-electron excess noise as computed in
our conserving kinetic model.
At low fields, where both
kinetic and phenomenological models behave quadratically
with $V$, the hot-electron noise is still dominant
\cite{fnl}.

We also model the effect of nonideal inelastic
scattering by plotting the second (upper-subband) noise
contribution as a function of three different collision-time
ratios $\zeta_2 = \tau_{{\rm in};2}/\tau_{{\rm el};2}$;
namely, $\zeta_2 = 0.6, 0.8$, and 1.
There is a pronounced loss in strength for the second
peak as the inelastic effects are made stronger.
The corresponding plots of conductance (right-hand scale)
are much less affected. The sharp falloff in the excess
thermal noise should therefore be
a prime signature of dynamical
processes that could modify ballistic transport as observed.

\end{multicols}

\begin{minipage}{16cm}

\begin{figure}[h]

%
\begin{center}
\includegraphics[width=0.40\textwidth]{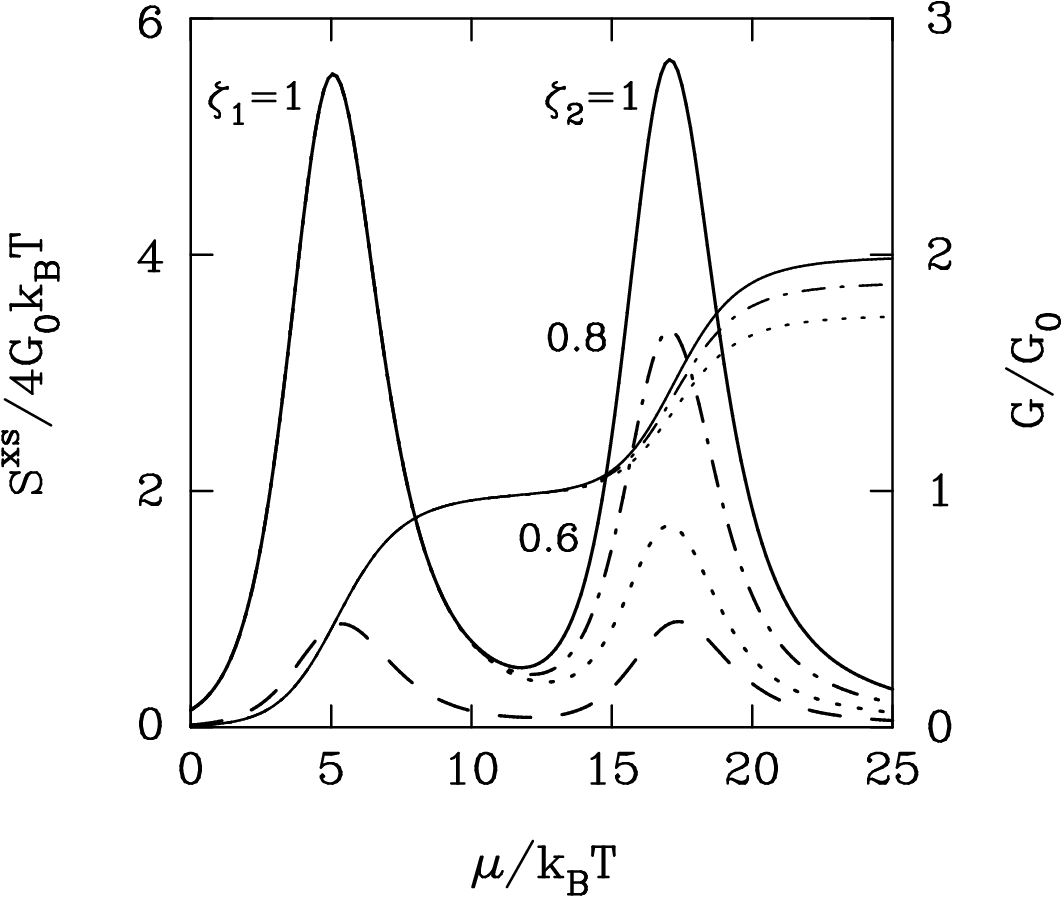}
\label{f5} 
%
\end{center}

{FIG. 5. Excess thermal noise and conductance
of a ballistic wire, calculated within a strictly
conserving kinetic model. Left scale: the excess noise
at the high voltage $V = 9k_{\rm B}T/e$,
normalized to the ideal ballistic Johnson-Nyquist noise
$4G_0k_{\rm B}T$,
is plotted as a function of chemical potential $\mu$.
Right scale: the corresponding quantized two-probe
conductance $G$, normalized to the universal quantum
$G_0 = e^2/\pi \hbar$. The large peaks in the excess noise
occur at the subband crossing points of $G$ located at energies
$\varepsilon_1 = 5k_{\rm B}T$ and $\varepsilon_2 = 17k_{\rm B}T$. 
The noise is remarkably high at the crossing points, where
the subband electrons are {\em classical}. It is low at the
plateaux in $G$, where subband {\em degeneracy} suppresses thermal noise.
There is a pronounced sensitivity of ${\cal S}^{\rm xs}$
to nonideality in $G$, controlled by
the ratio $\zeta_i = \tau_{{\rm in};i}/\tau_{{\rm el};i}$.
The smaller the ratio, the stronger the inelastic collisions.
${\cal S}^{\rm xs}$ manifests nonideality
much more strongly than $G$ itself.
Dashed line: the corresponding excess-noise prediction of
the nonconserving transmissive-diffusive theory;
see Eq. (\ref{16}).
It is much smaller than thermal hot-electron noise.
}

\end{figure}
%
\end{minipage}

\begin{multicols}{2}

\section*{5. Summary}

In this presentation we have stressed one idea above all:
that transport and noise are deeply intertwined. Their connection
is microscopic. This means that a microscopic
analysis (provided, for instance, by kinetic theory) is the
only effective vehicle for accessing
the physics of mesoscopic noise and
transport, in a logically seamless way.

There exists a distinctive set of fundamental identities that
{\em must} be satisfied within every truly microscopic model
of mesoscopic conduction.
The fluctuation-dissipation theorem is one such
\cite{nvk,kubo}.
It is essential to the understanding of noise as a phenomenon
conjoint with transport.

Alongside that basic theorem, the Fermi-liquid structure of the
electron gas provides the remaining fundamental relations: the
{\em sum rules}
\cite{pinoz}.
They are as critical to mesoscopic transport
as the fluctuation-dissipation relation itself. How scant
the regard has been for the electron-gas sum rules
within mesoscopics -- despite those rules'
long and thoroughly documented history
\cite{nozieres,pinoz,abri} --
can be gauged by the absence of any reference
to them, even in the most authoritative
accounts of contemporary mesoscopic theory
\cite{imry2,ferry,datta,blbu}.

Satisfaction of the sum rules is mandatory for any theory
that claims to describe degenerate electrons. This applies
most especially to every candidate model of mesoscopic noise.

In the area of nonequilibrium mesoscopic conduction,
we have covered the physical genesis and significance of
one of the primary sum rules, that for the
compressibility. There are three conclusions:

\begin{itemize}
\item
A correctly formulated kinetic theory of mesoscopic transport and
fluctuations, for {\em open} metallic conductors, will satisfy the
compressibility sum rule. This severely constrains the fluctuation
spectrum even at high fields. We have shown that the same, invariant,
sum rule is valid well beyond the near-equilibrium regime.

\item
In an {\em inhomogeneous} metallic conductor, strong internal Coulomb
correlations modify the fluctuations. They, and hence the
current noise, are self-consistently suppressed by the
increased electrostatic energy. The additional Coulomb
suppression lowers the value of the equilibrium compressibility.

The suppressed compressibility persists, without any alteration,
even when the degenerate system is driven out of equilibrium.
We predict that the signature of this suppression will be
found in reduced levels of excess hot-electron noise for certain
quantum-well-confined channels
\cite{gdii}.

\item
The compressibility sum rule is
{\em violated} by all mesoscopic noise models based on the
paradigm of (coherent) transmission linked to diffusion. The
latter, especially, is incompatible with the open reservoirs'
crucial function in controlling the magnitude of nonequilibrium
noise in a degenerate mesoscopic conductor.

The overall
{\em temperature scaling} of the thermal fluctuation spectrum
is a necessary consequence of degeneracy, expressed through
the compressibility sum rule. That scaling too
is violated by every transmissive-diffusive model, without exception.

\end{itemize}

Sum-rule violations place a prodigious question mark over a
theory's physical coherence. No amount of rationalization
can undo this degree of inconsistency.

In one dimension, our strictly conserving
kinetic theory of transport and noise
recovers -- as it should -- the quantized
Landauer conductance steps observed in open
(thus phase-incoherent) contacts
\cite{fnl}.
It also makes possible the calculation
of nonequilibrium hot-electron noise in a
one-dimensional ballistic device
\cite{fnl,balan,cmt25}.

As the carrier density in the device changes, striking peaks
appear in the excess thermal noise. These features contain detailed
information on the dynamics of nonideal transport in the sample.
They are {\em unrelated to shot noise}, which is a quite distinct
form of nonequilibrium electron-hole fluctuation.
Numerically, they dominate the
corresponding prediction of transmissive-diffusive phenomenology.

Elsewhere we apply our kinetic analysis of ballistic noise
to the celebrated quantum-point-contact noise
measurements by Reznikov {\em et al.}
\cite{rez}.
Our conservative kinetic computation shows that the
linear dispersion of excess current noise, with EMF,
is far from being the unique signature of shot noise.
The much-enhanced sensitivity of hot-electron noise to
electron-phonon processes, as we have discussed,
accounts for the observations equally well
\cite{balan,thak}.

In the future, we will expand our set of applications
to cover the fine details of low-dimensional
mesoscopic conduction. As to the Reznikov {\em et al.} data
\cite{rez},
a second and baffling set of observations should be examined:
the anomalous sequence of strong noise peaks at the
lowest subband threshold,
for fixed levels of the source-drain {\em current}.
There, the Landauer-B\"uttiker noise theory
\cite{blbu}
predicts, not the strong (and quite unexpected) peaks
that actually appear
\cite{rez},
but a totally featureless monotonic drop in the noise signal
right across the lowest subband threshold.

Those anomalous peaks have been analyzed
\cite{cmt25,thak}.
They are quite thermal. They respond in a most remarkable way
to field-induced, inelastic electron-phonon scattering.
Their resolution rests with the unexpected behavior
that kinetic theory reveals for the spectrum
of excited ballistic electrons.

\section*{Acknowledgments}

We thank Jagdish Thakur for agreeing to the use of
data from our collaboration, for his initiative in
extending our ballistic theory, and not least for
many insightful discussions. We also thank
Ann Osborne for valuable help with typescript preparation.

\end{multicols}

\end{document}